\documentclass[reprint,prl,aps,superscriptaddress]{revtex4-2}
\AtBeginDocument{\RenewCommandCopy\qty\SI}
\usepackage{times}
\usepackage{graphicx}
\usepackage{amsmath, braket, amsfonts}
\usepackage{amssymb}
\usepackage{sidecap}
\usepackage{bm, color, ulem}
\usepackage{xcolor}
\usepackage{ragged2e}
\usepackage{wrapfig,lipsum,booktabs}
\usepackage{lineno}
\usepackage{placeins}
\usepackage{makecell}
\usepackage{verbatim} % jack
\UseRawInputEncoding
\usepackage[utf8]{inputenc}

\usepackage{siunitx}

\newcommand{\be}{\begin{equation}}
\newcommand{\ee}{\end{equation}}
\usepackage{mathtools}

\DeclareUnicodeCharacter{03BD}{{$\nu$}}

\newcommand{\UCSB}{Department of Physics, University of California at Santa Barbara, Santa Barbara, California 93106, USA}

\makeatletter
\def\maketitle{
\@author@finish
\title@column\titleblock@produce
\suppressfloats[t]}
\makeatother

\AtBeginDocument{\RenewCommandCopy\qty\SI}

\begin{document}
\title{Cryogenic shock exfoliation for ultrahigh mobility rhombohedral graphite nanoelectronics}
\author{Ludwig Holleis}
\affiliation{\UCSB}
\author{Youngjoon Choi}
\affiliation{\UCSB}
\author{Canxun Zhang}
\affiliation{\UCSB}
\author{Jack H. Farrell}
\affiliation{Department of Physics and Center for Theory of Quantum Matter, University of Colorado, Boulder, Colorado 80309, USA}
\author{Gabriel Bargas}
\affiliation{\UCSB}
\author{Audrey Hsu}
\affiliation{\UCSB}
\author{Zexing Chen}
\affiliation{\UCSB}
\author{Ian Sackin}
\affiliation{\UCSB}
\author{Wenjie Zhou}
\affiliation{\UCSB}
\author{Yi Guo}
\affiliation{\UCSB}
\author{Thibault Charpentier}
\affiliation{\UCSB}
\author{Yifan Jiang}
\affiliation{\UCSB}
\author{Benjamin A. Foutty}
\affiliation{\UCSB}
\author{Aidan Keough}
\affiliation{\UCSB}
\author{Martin E. Huber}
\affiliation{Departments of Physics and Electrical Engineering, University of Colorado, Denver, Colorado 80204, USA}
\author{Takashi Taniguchi}
\affiliation{International Center for Materials Nanoarchitectonics,
National Institute for Materials Science,  1-1 Namiki, Tsukuba 305-0044, Japan}
\author{Kenji Watanabe}
\affiliation{Research Center for Functional Materials,
National Institute for Materials Science, 1-1 Namiki, Tsukuba 305-0044, Japan}
\author{Andrew  Lucas}
\affiliation{Department of Physics and Center for Theory of Quantum Matter, University of Colorado, Boulder, Colorado 80309, USA}
\author{Andrea F. Young}
\email{andrea@physics.ucsb.edu}
\affiliation{\UCSB}
\date{\today}

\begin{abstract}
{Rhombohedral multilayer graphene (RMG) offers a highly tunable platform for correlated
electron physics, featuring field-effect control of magnetic, superconducting, and topological
phases\cite{shi_electronic_2020,zhou_superconductivity_2021,zhou_half_2021,zhou_isospin_2022,chen_Evidence_2019,chen_Signatures_2019,arp_intervalley_2024,han_orbital_2023,han_large_2024,liu_spontaneous_2024,zhou_layerpolarized_2024,lu_fractional_2024,patterson_superconductivity_2025,choi_superconductivity_2025,han_signatures_2025,holleis_fluctuating_2025,auerbach_isospin_2025,zhang_layerdependent_2025,xie_tunable_2025,guo_flat_2025,kumar_superconductivity_2025,qin_extreme_2025,xie_MagneticFieldDriven_,zhang_Imaging_2026a,deng_magneticfieldinduced_2026,morissette_evidence_2025,sheekey_visualizing_2026}.
The promise of these materials has been held back by the limited abundance of rhombohedral
stacking in natural graphite, which constrains both sample yield and useful area. Here we introduce `cryogenic shock exfoliation' to produce large-area rhombohedral graphene flakes which, combined with a low-pressure van der Waals assembly technique that preserves
stacking order, enable highly uniform devices exceeding $1300\,\mu\mathrm{m}^2$ with fabrication yields of ${\sim}90\%$. Using scanning nanoSQUID-on-tip imaging, we demonstrate uniform spin magnetism over the full central $10\times10$~$\mu$m$^2$ area
of our devices. Transverse magnetic focusing reveals a disorder mean free path exceeding $200\,\mu$m at low temperatures. Within the flat surface bands of RMG\cite{guo_flat_2025}, we observe a size-driven crossover from Poiseuille to porous electron flow in the intermediate-temperature regime of strong electron-electron hydrodynamics\cite{zhang_Imaging_2026,holleis_fluctuating_2025}, providing a further signature of ultrahigh device quality. Our approach overcomes a key materials bottleneck in the fabrication of mesoscopic rhombohedral graphene devices, paving the way for incorporating strongly correlated phases into two-dimensional nanoelectronics.}
\end{abstract}
\maketitle 

\section{Introduction}
The isolation of graphene demonstrated that mechanically cleaving graphite could isolate atomically thin crystals with electronic properties inaccessible in any bulk form\cite{novoselov_electric_2004}. 
The same technique has since been applied to hundreds of layered materials and extended to the deliberate construction of van-der-Waals (vdW) heterostructures assembled layer by layer from exfoliated flakes\cite{geim_Van_2013}. However, mechanical exfoliation and van der Waals assembly provides little control over metastable configurations of the constituent crystals, a significant limitation in light of the fact that metastable phases often host the richest correlated electron phases. 
This phenomenon is illustrated most dramatically by graphene multilayers with rhombohedral stacking, where a dizzying array of magnetic, superconducting, and topological states have been observed at low carrier density\cite{shi_electronic_2020,zhou_superconductivity_2021,zhou_half_2021,zhou_isospin_2022,chen_Evidence_2019,chen_Signatures_2019,arp_intervalley_2024,han_orbital_2023,han_large_2024,liu_spontaneous_2024,zhou_layerpolarized_2024,lu_fractional_2024,patterson_superconductivity_2025,choi_superconductivity_2025,han_signatures_2025,holleis_fluctuating_2025,auerbach_isospin_2025,zhang_layerdependent_2025,xie_tunable_2025,guo_flat_2025,kumar_superconductivity_2025,qin_extreme_2025,xie_MagneticFieldDriven_,zhang_Imaging_2026a,deng_magneticfieldinduced_2026,morissette_evidence_2025,sheekey_visualizing_2026}. 

Among allotropes of carbon, graphite with Bernal stacking, in which layers are shifted by one third of a unit cell in an alternating pattern, is the lowest energy structure. 
Rhombohedral stacking, in which successive layers are shifted by a third of a unit cell in the same direction, is slightly higher energy.  
Nevertheless, natural graphite is found to have approximately 10--15\% fraction of rhombohedral stacking\cite{lipson_structure_1942}.  
A similar fraction of rhombohedral multilayer graphene is found in mechanically exfoliated flakes\cite{lui_imaging_2011}, with typical domain sizes rarely exceeding 100-200 $\mu m^2$. 
The small size and relative scarcity of rhombohedral domains severely limits the ultimate size and throughput of rhombohedral graphene devices.  
Yields are further compromised by the fact that that rhombohedral stacking converts readily to the Bernal configuration during the van der Waals assembly process. 
Despite these limitations, experiments have shown rhombohedral graphene multilayers to be an exceptionally low-disorder platform for correlated electron physics, with elastic mean-free path typically limited only by the sample size\cite{zhou_superconductivity_2021} and high experimental reproducibility across devices.  
These findings motivate experimental efforts to address the bottleneck represented by the scarcity of rhombohedral graphene source material.  

\begin{figure*}
    \centering
    \includegraphics[width=\textwidth]{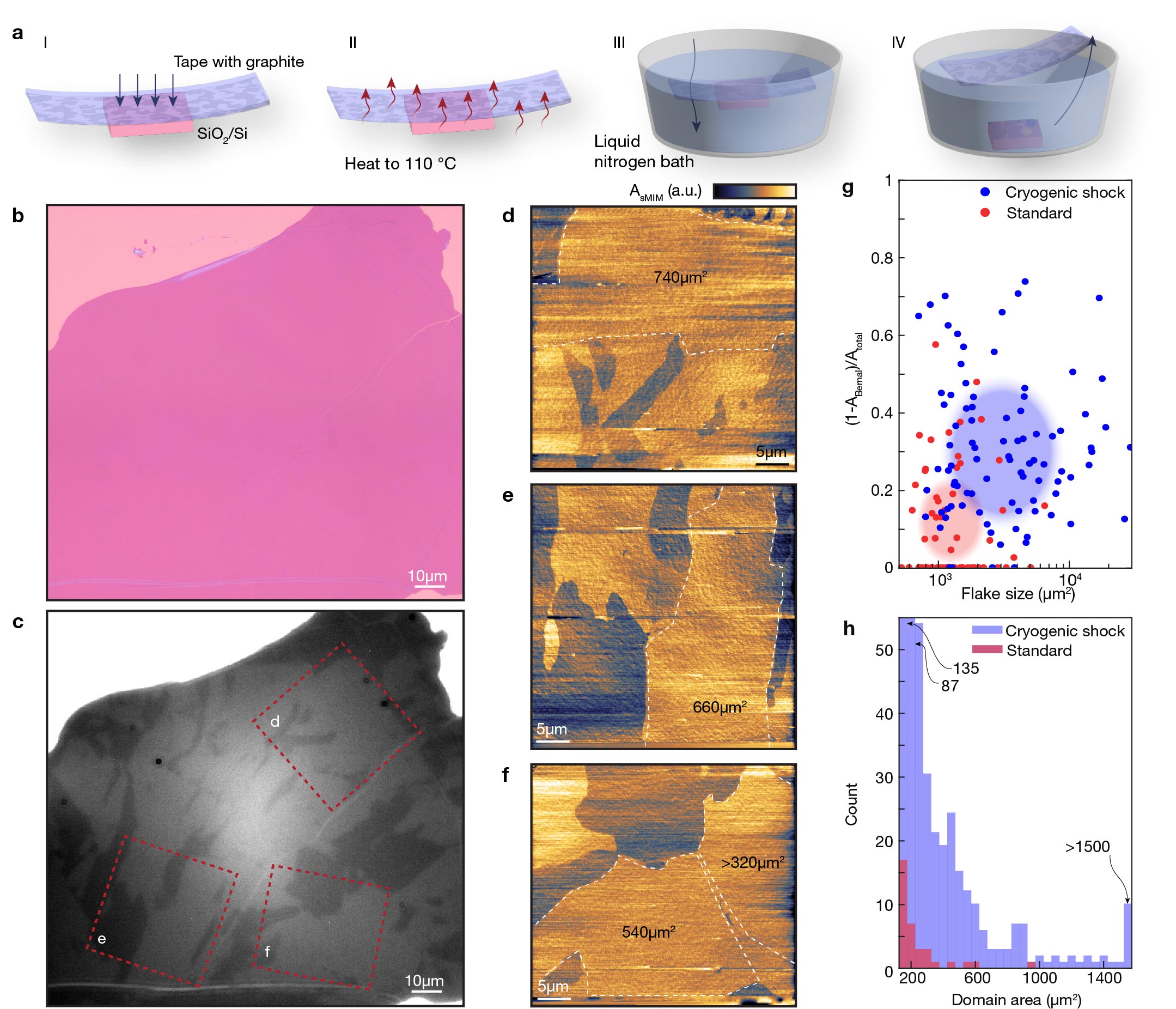}
    \caption{\textbf{Cryogenic shock exfoliation of rhombohedral graphene.}
    \textbf{(a)} Step-by-step illustration of the cryogenic shock exfoliation process.
    \textbf{(b)} Optical micrograph taken in the visible and \textbf{(c)} infrared.  
    Bright regions in the infrared image correspond to rhombohedral stacking order of this $\sim$14 layer RMG flake.
    \textbf{(d-f)} Scanning microwave impedance microscopy images in the regions  marked in panel c.  
    Area of select contiguous rhombohedral domains are indicated. 
    \textbf{(g)} Total flake area versus non-Bernal stacking fraction for multilayer graphene flakes. 
    Shaded ovals are centered on the geometric mean for x-axis and arithmetic mean for the y-axis, and spanning one standard deviation.
    \textbf{(h)} Distribution of single-stacking-order, non-Bernal domain areas. See additional information in Methods and Fig.~\ref{figS_exfoliation_statistics}.}
    \label{fig:1}
\end{figure*}

Here, we introduce cryogenic shock exfoliation as an easily implementable approach to selectively increase the yield of metastable stacking orders. 
Prior work has shown that the energetic competition between rhombohedral and Bernal stacking is closely balanced, with rhombohedral stacking selectively favored by the application of electric fields or local mechanical forces\cite{yankowitz_electric_2014,jiang_Manipulation_2018,yang_stacking_2019,li_Global_2020,gao_largearea_2020,bouhafs_synthesis_2021,dey_Uniaxial_2024}.  
Our technique is inspired by numerical simulations which suggest that the energetic competition between rhombohedral and Bernal stacking can be tuned using shear stress\cite{nery_LongRange_2020}.  
We induce a large stress in graphite flakes during the exfoliation process by leveraging the differential thermal contraction between polymer based tape, graphite, and a silicon substrate upon rapid immersion in liquid nitrogen. 
We find that flakes produced by this `cryogenic shock' method contain ${\sim}32\%$ rhombohedral stacking, several times more than in flakes produced by typical methods.  
More importantly, we routinely identify uniform rhombohedral domains exceeding $1000\,\mu\mathrm{m}^2$. 
Combined with a low-pressure vdW-assembly technique that suppresses stacking-order relaxation during transfer, we achieve fabrication yields of ${\sim}90\%$ and demonstrate rhombohedral devices with areas exceeding $1300\,\mu\mathrm{m}^2$ - making our methods the first to reliably fabricate strongly correlated vdW-devices based on metastable materials, including moir\'e systems.
We establish the quality of these devices through three independent
probes: nanoSQUID-on-tip imaging reveals uniform spin magnetization over the entire $10\times10\,\mu\mathrm{m}^2$ device area; transverse magnetic focusing establishes a disorder mean free path exceeding $200\,\mu\mathrm{m}$ at low temperature; and temperature-dependent
transport in channels of varying width reveals a crossover from Poiseuille to porous electron flow, a hydrodynamic regime accessible only when the device dimension substantially exceeds the Gurzhi length\cite{gurzhi_MINIMUM_1963}. 
Our results establish that strongly correlated phases in a metastable vdW material can be purposefully stabilized on length scales compatible with a wide range of mesoscopic quantum electronics.

\begin{figure*}
    \centering
    \includegraphics[width=\textwidth]{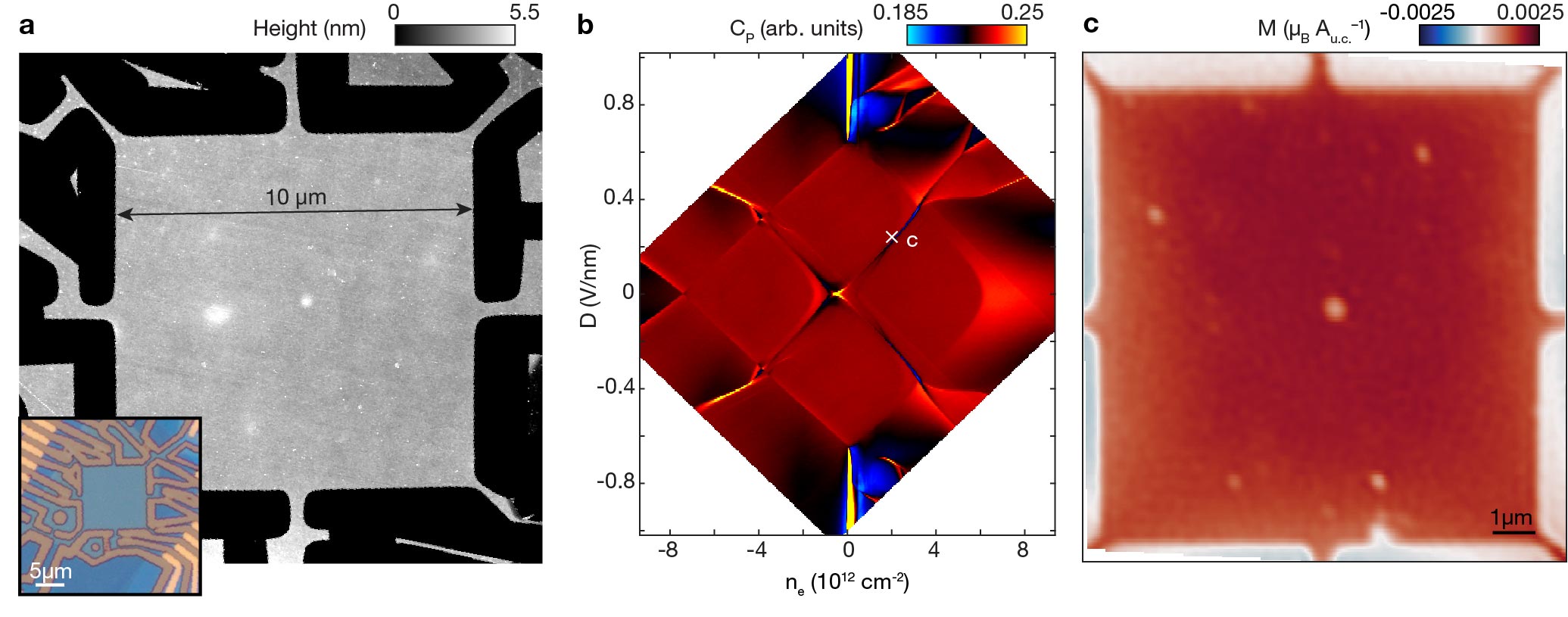}
    \caption{\textbf{Real-space characterization of device uniformity.}
    \textbf{(a)} Atomic force micrograph of a 13-layer dual-gated RMG device. Inset: optical micrograph.
    \textbf{(b)} Penetration field capacitance $C_P$ measured across the accessible range of $n_e$ and $D$.
    \textbf{(c)} Real space magnetization $M$ determined by local magnetometry.  A fringe magnetic field image is acquired by modulating the back gate across a phase boundary separating a spin-polarized half-metal (point `X' in panel b) and a non-magnetic phase.  Standard Fourier-domain algorithms are then used to reconstruct $M$ from the measured fringe fields (see Methods).}
    \label{fig:2}
\end{figure*}

\section{Cryogenic shock exfoliation}

The cryogenic shock exfoliation process is illustrated in Fig.~\ref{fig:1}a. Bulk graphite crystals are mechanically cleaved with tape and pressed onto a silicon substrate coated with  $285\,\mathrm{nm}$ of thermally grown SiO$_2$. 
Both tape and chip are then heated to $110\,^\circ$C, and then together fully submerged in liquid nitrogen to achieve a large thermal differential via the cryogenic shock.  
As both cool rapidly, a large stress gradient develops between tape and chip and the tape detaches from the silicon chip. 
After thermal equilibration in LN$_2$, the chip is either immediately reheated, or slowly warmed in a dry atmosphere to room temperature to avoid condensation of water on the freshly exfoliated graphene surface.

Figures~\ref{fig:1}b and~\ref{fig:1}c show optical micrographs of a representative flake in the visible and infrared spectral range, respectively. 
While the visible image reveals a flake of uniform thickness, the infrared contrast is sensitive to the stacking order\cite{lu_extended_2025,feng_rapid_2025} and shows bright regions identified as rhombohedral stacking and darker regions identified with Bernal stacking. 
Scanning microwave impedance microscopy (sMIM)\cite{holleis_nanoscale_2025} images at three representative locations (Fig.~\ref{fig:1}d-f) confirm uniformity down to the nano-meter scale of the crystal structure over hundreds of $\mu$m$^2$. 
The flake shown is a prototypical example; additional examples of different thicknesses are shown in Fig.~\ref{figS_additional_flakes}, illustrating that uniform rhombohedral areas exceeding $1000\,\mu$m$^2$ are routine. 

To quantify the improvement over standard exfoliation, we compare both methods on equal substrate areas (see Methods and Fig.~\ref{figS_exfoliation_statistics}). 
The distributions are shown in Fig.~\ref{fig:1}g: cryogenic shock exfoliation increases both the total flake area and the rhombohedral fraction, with averages rising from $1200\,\mu$m$^2$ and $12\%$  to $3100\,\mu$m$^2$ and $32\%$. 
The standard technique values agree with previously reported crystallographic ratios in both exfoliated flakes\cite{lui_imaging_2011} and bulk graphite\cite{lipson_structure_1942}, suggesting that the cryogenic shock induces a partial conversion of Bernal to rhombohedral stacking, likely via differential thermal contraction induced stress\cite{laves_Formation_1956,yang_stacking_2019,nery_LongRange_2020,dey_Uniaxial_2024}.
Figure~\ref{fig:1}h quantifies the distribution of contiguous rhombohedral domain sizes.  
In addition to a significant increase in the number of domains with area $ <500\,\mu$m$^2$, cryogenic shock exfoliation produces a long tail of uniform rhombohedral domains beyond  $\sim 1500\,\mu$m$^2$ that are never found using conventional techniques - independent of layer number (see Fig.~\ref{figS_exfoliation_statistics}).

Leveraging the large flake areas produced by cryogenic shock exfoliation requires preserving the stacking order through the van der Waals assembly process. 
Prior work has shown the importance of isolating pure rhombohedral domains by cutting prior to transfer\cite{chen_Evidence_2019,zhou_half_2021},which prevents structural relaxation via domain wall motion. To this end, we use scanning microwave impedance microscopy to ensure the absence of sub-diffraction limit Bernal domains that may serve as nucleation sites\cite{holleis_nanoscale_2025}.  
Performing the lamination procedure along crystallographic directions has also bee shown to suppress relaxation\cite{yang_stacking_2019}. 
We further introduce a low-pressure transfer technique in which a polycarbonate (PC) film is suspended over a PDMS micro-cavity (see Fig.~\ref{figS_pressureless_pickup}a). 
The transfer utilizing a membrane\cite{wang_clean_2023,wu_SO5_2014} applies minimal pressure to the RMG flake during both initial pickup (via a hexagonal boron nitride flake) and during subsequent pickup of hexagonal Boron Nitride and graphite flakes. 
Compared to conventional methods\cite{zhou_half_2021}, low-pressure transfer increases RMG yield from ${\sim}30\%$ to $>80\%$ with statistics acquired over a sample of 24 and 62 devices, respectively (see Fig.~\ref{figS_pressureless_pickup}b). 
Combining all known techniques to reduce relaxation (such as imaging and isolation of poor rhombohedral domains and transfer along crystallographic directions) produces a final device yield of ${\sim}90\%$ (see Methods and Fig.~\ref{figS_pressureless_pickup}).

\begin{figure*}
    \centering
    \includegraphics[width=\textwidth]{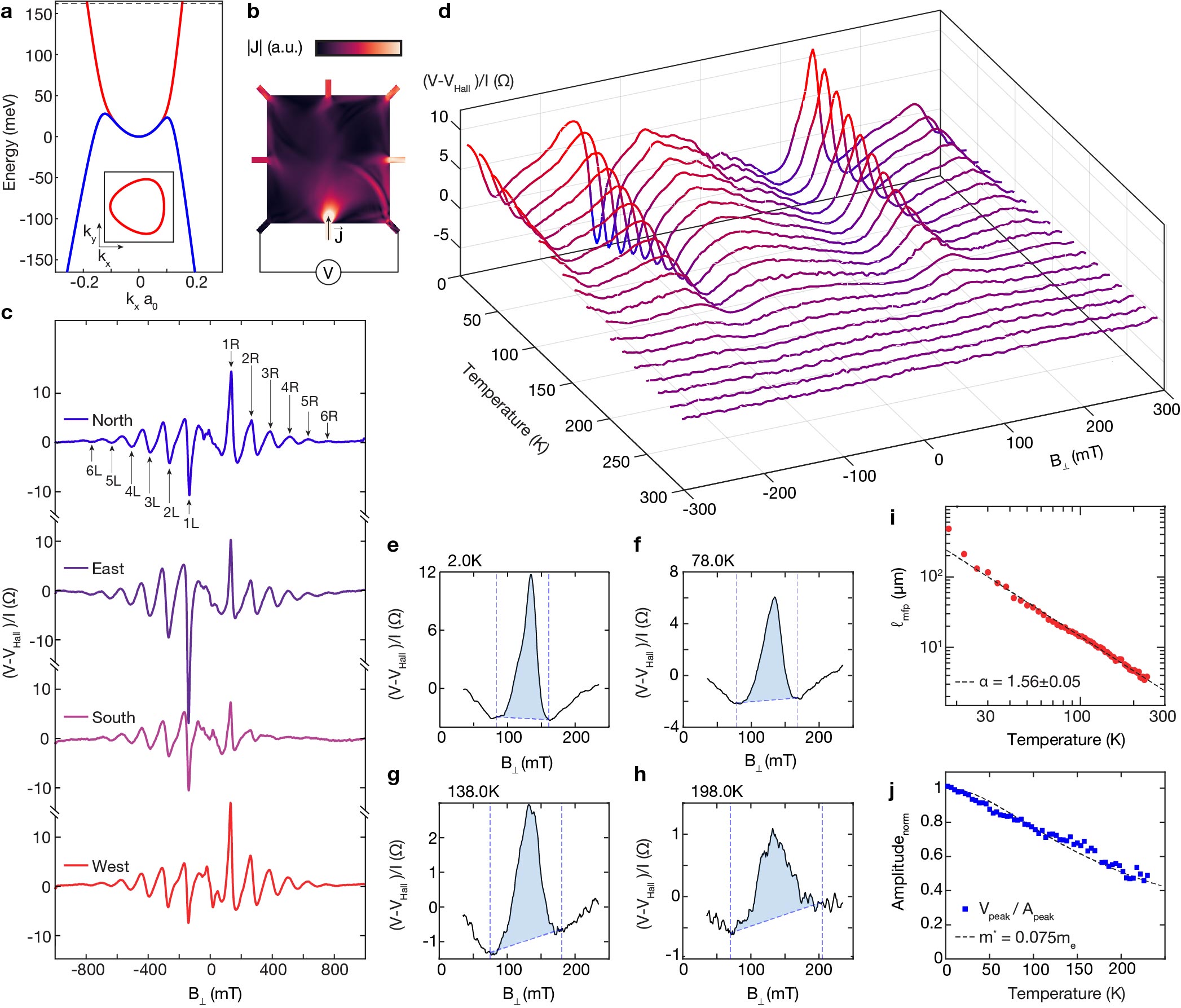}
    \caption{\textbf{Transverse magnetic focusing.}
    \textbf{(a)} Single-particle band structure of rhombohedral 13-layer graphene at $D=0$. The inset shows the Fermi surface at the dashed line ($E_F = 160\,\mathrm{meV}$),
    corresponding to $n_e \approx 10^{13}\,\mathrm{cm}^{-2}$. 
    \textbf{(b)} Simulations of the current magnitude $|J|$ for a fully ballistic Boltzmann model with applied magnetic field $B=135$~mT, corresponding to the first TMF resonance.
    \textbf{(c)} TMF data at $n_e = 10^{13}\,\mathrm{cm}^{-2}$ for four contact configurations along the different edges of the devices. 
    \textbf{(d)} TMF data for the `South' edge as a function of temperature.
    \textbf{(e)} First TMF peak for at T=2.0K, 
    \textbf{(f)} 78.0K, 
    \textbf{(g)} 138.0K, and 
    \textbf{(h)} 198.0K. 
    The blue shaded region is the peak area $A_\mathrm{peak}$.
    \textbf{(i)} Temperature-dependence of $\ell_\mathrm{mfp}$ extracted from the TMF data. 
    The dashed line is a fit to Matthiessen's rule (see main text). 
    \textbf{(j)} Temperature-dependence of  $V_\mathrm{peak}/A_\mathrm{peak}$, normalized to the low-temperature limit. The dashed line is a single-parameter fit to a thermal broadening model for effective mass $m^*=.075 m_e$.}
    \label{fig:3}
\end{figure*}

\section{Characterizing sample uniformity}

Fig.~\ref{fig:2}a shows optical and atomic force micrographs of a dual-graphite gated 13-layer RMG device fabricated from a cryogenic shock exfoliated flake using the methods of the previous section. 
The total rhombohedral area is $740\,\mu$m$^2$, exceeding maximum sizes of rhombohedral devices in the literature by at least a factor of two.  Additional devices with areas exceeding $1300\,\mu$m$^2$ are shown in Fig.~\ref{figS_devices}. 
Fig.~\ref{fig:2}b shows the penetration field capacitance as a function of the charge carrier density $n_\mathrm{e}$ and displacement field $D$. 
The phase diagram shows numerous features known to be associated with spin and valley ordering\cite{guo_flat_2025}. 
Notably, no additional features are observed.
Given that the capacitance measurement averages the signal over the entire sample area, this suggests uniform stacking across the entire dual-gated region.
Data from additional, thinner RMG devices prepared via cryogenic shock exfoliation can be found in Fig.\ref{figS_hexalayer} and reference\cite{sheekey_visualizing_2026}, showing superconductivity and symmetry broken phases equivalent to results on devices produced with standard techniques reported in the literature\cite{choi_superconductivity_2025,xie_MagneticFieldDriven_,morissette_evidence_2025,deng_magneticfieldinduced_2026}.
To confirm the sample uniformity, we use nanoSQUID-on-tip (nSOT) magnetometry to image the magnetization in real space.  Figure~\ref{fig:2}c shows the out-of-plane magnetization within the spin-polarized half-metal, measured at the $(n_\mathrm{e}, D)$ point marked in Fig.~\ref{fig:2}b.
The spatial scan reveals uniform magnetization over the entire square, with only small magnetically inactive regions correlated with hydrocarbon residues trapped between the graphene and dielectric layers, most of which are visible in the atomic force micrograph.

We also characterize the transport behavior using transverse magnetic focusing (TMF)\cite{vanhouten_Coherent_1989,taychatanapat_Electrically_2013}. 
We focus on the high electron density regime, $n_e = 10^{13}\,\mathrm{cm}^{-2}$, where only a single, low-effective-mass band is occupied (see Fig.~\ref{fig:3}a). 
In TMF, current is injected through a contact on the edge of the $10\times10\,\mu$m$^2$ square and focused by an out-of-plane magnetic field (see Fig.~\ref{fig:3}b), producing a peak in the measured voltage between the adjacent corner contacts whenever $B = 2\hbar k_F/(ed)$, where  $k_F$ the Fermi wave vector and $d \approx 5\,\mu$m is the contact separation. 
Figure~\ref{fig:3}c shows measurements of this voltage differential (after subtracting a linear-in-B Hall voltage $V_{Hall}$) measured on all four sides of the device. All four configurations display similar behavior with up to six observable peaks, corresponding to between zero and 5 boundary reflections. Qualitatively, these results imply a mean free path at least comparable to the device scale, and high degree of specularity to the boundary scattering.  

Tracking the focusing peaks as a function of increasing temperature (Fig.~\ref{fig:3}d-h) reveals a gradual suppression of peak amplitude.  This can be attributed to a combination of thermal broadening of the occupation and increased scattering. 
To separate these contributions, we measure both the height ($V_\mathrm{peak}$) and area ($A_\mathrm{peak}$) of the voltage peaks.  
When $k_\mathrm{B}T/E_F < w/d$ (here $w$ is the collector width), $A_{peak}$ is sensitive only to $\ell_\mathrm{mfp}$\cite{ingla-aynes_Specular_2023a}. 
The resulting $\ell_\mathrm{mfp}(T)$, shown in Fig.~\ref{fig:3}i, is well described by Matthiessen's rule $\frac{1}{\ell_\mathrm{mfp}(T)} = \frac{1}{\ell_0} + C\,T^\alpha$, with $\alpha = 1.56\pm0.05$. 
We find $\ell_\mathrm{mfp} > 200\,\mu$m below 30~K, extrapolating to  $\ell_\mathrm{mfp}(300\,\mathrm{K}) \approx 2\,\mu$m at room temperature. 
This is comparable to the mean free paths reported for the best-in-class mono- and bi-layer  graphene devices\cite{mayorov_Micrometerscale_2011,wang_Onedimensional_2013,wang_clean_2023}.
Notably, negative resistances associated with ballistic electron transport are also observed  when $\ell_\mathrm{mr}$ exceeds the device size without applying a magnetic field (see Fig.~\ref{figS_negativeR}).
$V_\mathrm{peak}$, in contrast, includes a contribution from the thermally broadened Fermi distribution, and is consequently sensitive to both the mean free path and effective mass.  
The ratio $V_\mathrm{peak}/A_\mathrm{peak}$ isolates the thermal broadening contribution, which we find to be well fit by a simple model in which the effective mass is the only adjustable parameter.  The best fit of $m^* = 0.075 \pm 0.05\,m_e$ (Fig.~\ref{fig:3}j) is in excellent agreement with $m^* = 0.078\,m_e$ determined from a tight-binding band structure model (see Methods).

\begin{figure*}
    \centering
    \includegraphics[width=\textwidth]{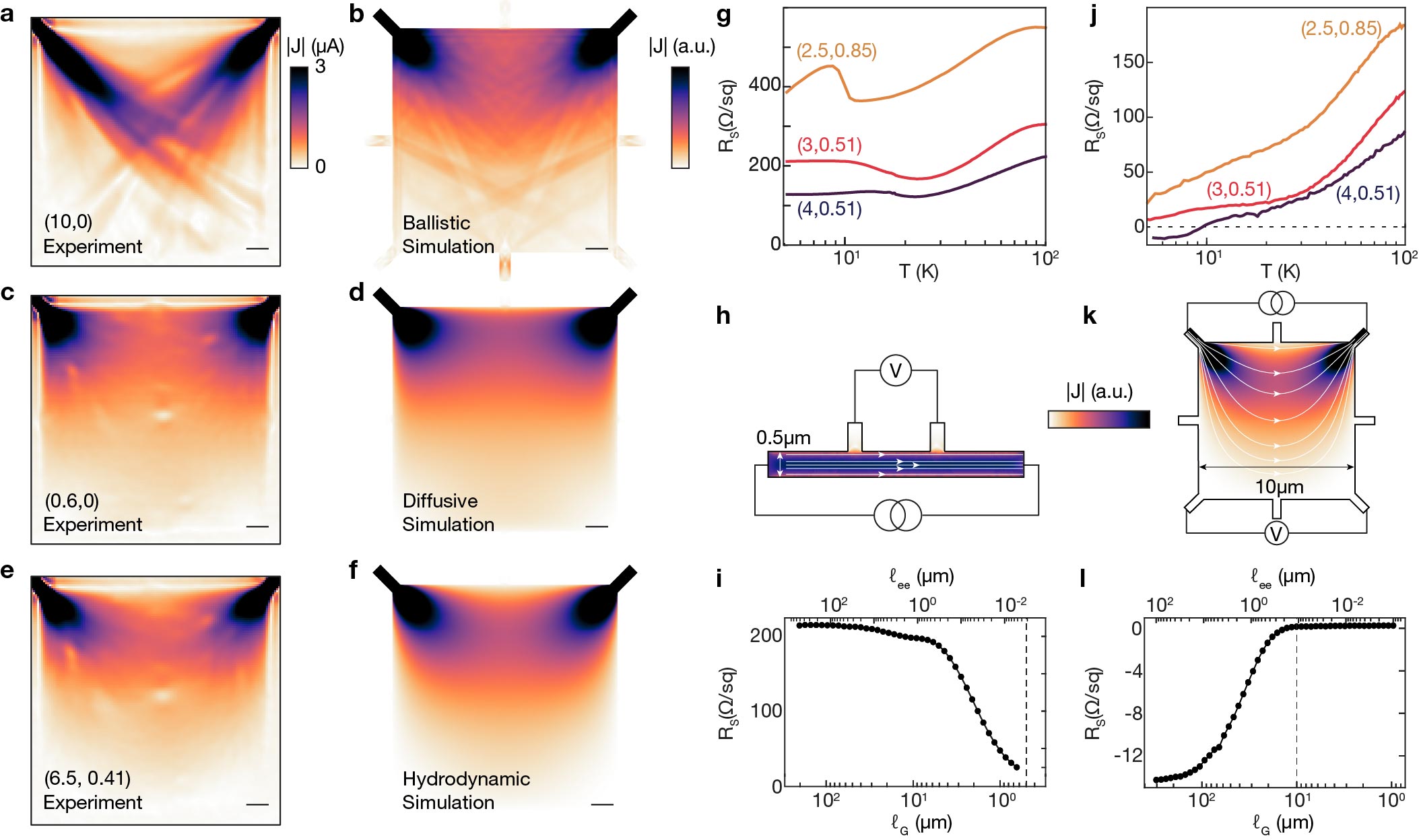}
    \caption{\textbf{Ballistic, diffusive, and hydrodynamic electron transport regimes in RMG.}
    \textbf{(a)} nSOT Biot--Savart imaging of $10\,\mu$A current flow at $n_e = 10^{13}\,\mathrm{cm}^{-2}$, $D = 0\,\mathrm{V/nm}$.
    \textbf{(b)} Simulation of current flow in the ballistic regime ($\ell_\mathrm{ee} = 200\,\mu$m, $\ell_\mathrm{mr} = 200\,\mu$m), obtained by solving the quantum kinetic equation (see Methods).
    \textbf{(c)} nSOT image at $n_e = 0.6\times10^{12}\,\mathrm{cm}^{-2}$, $D = 0\,\mathrm{V/nm}$, close to charge neutrality.
    \textbf{(d)} Simulation in the diffusive regime ($\ell_\mathrm{ee} = 50\,\mathrm{nm}$, $\ell_\mathrm{mr} = 10\,\mu$m).
    \textbf{(e)} nSOT image at $n_e = 6.5\times10^{12}\,\mathrm{cm}^{-2}$, $D = 0.41\,\mathrm{V/nm}$, in the spin- and valley-unpolarized single-surface regime.
    \textbf{(f)} Simulation in the hydrodynamic regime ($\ell_\mathrm{ee} = 50\,\mathrm{nm}$, $\ell_\mathrm{mr} = 200\,\mu$m).
    Scale bar: $1\,\mu$m for all images.
    \textbf{(g)} Temperature-dependent sheet resistance in the narrow channel. 
    The orange, red and purple curves are taken at the ($n_\mathrm{e}$, $D$) (in units [$10^{12}\,\mathrm{cm}^{-2}$,$\mathrm{V/nm}$]) in the single-surface regime of 13-layer RMG.
    \textbf{(h)} Simulation of hydrodynamic current flow through a narrow $0.5\,\mu$m channel with $\ell_\mathrm{ee} \approx 50\,\mathrm{nm}$ and $\ell_\mathrm{mr} = 200\,\mu$m.
    \textbf{(i)} Simulated channel resistance as a function of the Gurzhi length $\ell_\mathrm{G} = \sqrt{\ell_\mathrm{ee}\ell_\mathrm{mr}}$ with $\ell_\mathrm{mr} = 200\,\mu$m fixed. 
    The dashed line marks the channel width.
    \textbf{(j)} Temperature-dependent resistance in the large square under the same conditions as panel h.
    \textbf{(k)} identical simulation as panel f, including the contact configuration.
    \textbf{(l)} Simulated resistance of the large square as a function of $\ell_\mathrm{G}$. 
    A full resistance map as a function of $\ell_\mathrm{mr}$ and $\ell_\mathrm{ee}$ is shown in Fig.~\ref{figS_simulations}.}
    \label{fig:4}
\end{figure*}

\section{Ballistic, diffusive, and hydrodynamic current flow in RMG}

When the disorder mean free path is much longer than the device size $L$, the nature of electronic transport is dictated by the interplay of electron-electron and electron-hole collisions - enabling access to non-diffusive transport regime. 
The ballistic transport regime described in Figure \ref{fig:3} occurs when both processes are weak, so that electrons scatter only over distances much larger than the device dimensions. However, when these processes become stronger, charge transport is expected to become either diffusive (when current relaxing electron-hole collisions dominate) or hydrodynamic (when current conserving electron-electron collisions dominate).  

To disambiguate these regimes, we directly image the Biot-Savart magnetic fields arising from the microscopic current flow pattern at a base temperature of $T\approx 1.6K$ and an applied current of 10~$\mu$A.
Fig.~\ref{fig:4}a shows the current density $|J|$ measured under similar conditions to Fig.~\ref{fig:3}, where bulk transport is consistent with the ballistic limit. The current flow profile shows sharply  defined current jets emanating from the contacts. 
This behavior is reproduced using numerical simulations of the quantum kinetic equation\cite{farrell_Characterizing_2026} in the ballistic limit, where both the current conserving collision length $\ell_{ee}$ and current relaxing collision lengths $\ell_{mr}$ are much larger than $L$ (see Fig.~\ref{fig:4}b and Methods).
We attribute the quantitative difference of experiment and simulations to details of  the current injection at the contacts.
Qualitatively different behavior is observed near overall charge neutrality.
In this regime, the system is a nearly-compensated semi-metal, and transport is dominated by electron-hole and electron-electron collisions. 
The current jets disappear and the flow concentrates along the top sample boundary (Figs.~\ref{fig:4}c).
This is most consistent with $l_{ee}\approx 50nm$ and $\ell_{mr}\approx 10\mu m$, placing the sample on the edge of the diffusive regime, with significant momentum relaxation occurring on length scale of the device size (see also Fig.\ref{figS_hydro_diffusive}). 
Finally, Fig.~\ref{fig:4}e shows the current density at higher displacement field, where the electron system occupies a single, surface polarized band with a larger effective mass\cite{guo_flat_2025}. 
In this regime electron-electron interactions are expected to dominate while electron-hole scattering is absent.  We indeed observe a third distinct flow profile: the current is concentrated in the bulk, far from the sample boundary  (Figs.~\ref{fig:4}e). 
We attribute this to electron hydrodynamics\cite{lucas_Hydrodynamics_2018a}, and is consistent with Boltzmann simulations shown in Fig.~\ref{fig:4}f for 
$\ell_\mathrm{ee} = 50\,\mathrm{nm} \ll \ell_\mathrm{mr} = 200\,\mu$m. 
These results are in agreement with recent findings in bilayer  graphene\cite{zhang_Imaging_2026} and are only achievable in high quality devices where $\ell_\mathrm{mr}$ significantly exceeds their size.

The large device area enabled by cryogenic shock exfoliation also allow us to tune between electron flow regimes via control of sample dimensions. 
Figure~\ref{fig:4}g shows temperature dependent transport measured in a $0.5\,\mu$m-wide channel in the single-surface regime; the three curves correspond to values of $n_\mathrm{e}$ and $D$ where the ground state is either a half- or quarter-metal\cite{guo_flat_2025}. 
Both regimes are metallic, with $R<<h/e^2$, but feature a region where $dR/dT<0$ and a resistance minimum around  $15$-$30\,\mathrm{K}$. 
This phenomenon was observed previously in bi- and trilayer RMG\cite{holleis_fluctuating_2025} and identified with the Gurzhi effect\cite{gurzhi_MINIMUM_1963,lucas_Hydrodynamics_2018a,zhang_Imaging_2026}. 
As $\ell_\mathrm{ee}$ decreases with increasing $T$, the viscous boundary layer thins, reducing the momentum transfer between the electron fluid and the channel walls and thereby lowering the resistance.
Simulated $|J|$ in a channel geometry with similar values for $\ell_{ee}$ and $\ell_{mr}$ as for Fig.~\ref{fig:4}f is shown in Fig.~\ref{fig:4}h. 
The current distribution is depressed near the sample boundaries, consistent with Poiseuille flow\cite{sulpizio_Visualizing_2019, ku_Imaging_2020,Jenkins_Imaging_2022}.  
We may then estimate the temperature dependence by treating $\ell_\mathrm{ee}^{-1}$ as a proxy for the temperature.  
Fig.~\ref{fig:4}j shows the channel resistance per square, $R_s$, as a function of the Gurzhi length $\ell_G=\sqrt{\ell_{ee}\ell_{mr}}$, with  $\ell_\mathrm{mr}$ held fixed.  
$R_s$ decreases with decreasing Gurzhi length, consistent with the experiment. 

In contrast, in the $10\times10\,\mu$m$^2$ square geometry
(Fig.~\ref{fig:4}j), the resistance increases monotonically with temperature with no minimum at intermediate temperatures. 
Simulations show that boundary effects are much weaker in this geometry, with very little current flowing within $\ell_{G}$ of the boundary (Fig.~\ref{fig:4}k).  
As a result, the calculated resistance is nearly insensitive to $\ell_\mathrm{ee}$ when $\ell_{ee}<10\mu m$, as shown in Fig.~\ref{fig:4}l.  
Notably, the effect of decreasing $\ell_{ee}$ is opposite as well: at large values of $\ell_{ee}$, the simulated resistance is negative due to ballistic `bend resistance' effects that are absent in the channel geometry. This is consistent with experiments at the largest values of $n_e$ where electron-electron interactions are expected to be weakest (see also Fig. \ref{figS_negativeR}).  
$R_s(T)$ thus increases with increasing $T$ from its negative (or small) low temperature value.  
Contrasted to the Poiseuille regime, the electron fluid flows unconstrained through the bulk, relaxing momentum only through residual impurity and phonon scattering. 
This is the porous flow regime\cite{kumar_imaging_2022,stern_how_2022}, in which momentum is dissipated volumetrically rather than at the boundaries — a flow regime distinct from ohmic transport because electron-electron scattering continues to dominate ($\ell_\mathrm{ee} \ll \ell_\mathrm{mr}$). 
The crossover between Poiseuille and porous flow, achieved here simply by increasing the device width at fixed parameters, constitutes direct experimental access to a hydrodynamic phenomenon that requires simultaneously ultra-low disorder and device dimensions at which boundary effects are negligible — conditions that RMG at this scale of uniformity now satisfies.

\section{Conclusion}

We have introduced cryogenic shock exfoliation as a route to large-area, high-purity rhombohedral multilayer graphene, and demonstrated a van der Waals assembly technique that preserves metastable stacking order through the heterostructure fabrication process. 
Together these advances raise the standard metrics of RMG device quality---area, yield, and uniformity--into a qualitatively new regime. 
In the near term, these methods enable previously impractical experiments, including high spectral resolution ARPES or the creation of monolithic superconducting resonators and planar Josephson junctions based on intrinsic rhombohedral graphene superconductivity. 
More broadly, cryogenic shock exfoliation transforms RMG from a materials curiosity into a reproducible platform in which the rich many-body physics pioneered in metastable two-dimensional structures -- moiré and crystalline alike -- becomes scalable for van-der-Waals nanoelectronics.

\section*{Acknowledgements} 
Experimental work at UCSB was primarily supported by the Department of Energy under Award DE-SC0020043 to A.F.Y.
Additional funding for nSOT sensor and microscope development was provided by the Experimental Physics Investigator program of the Gordon and Betty Moore Foundation under Award GMBF-13801 to A.F.Y. 
J.H.F. and A.L. are supported by the National Science Foundation under CAREER Grant DMR-2145544.
This work utilized the Alpine high performance computing resource at the University of Colorado Boulder. 
Alpine is jointly funded by the University of Colorado Boulder, the University of Colorado Anschutz, Colorado State University, and the National Science Foundation (award 2201538).
K.W. and T.T. acknowledge support from the JSPS KAKENHI (Grant Numbers 21H05233 and 23H02052), the CREST (JPMJCR24A5), JST and World Premier International Research Center Initiative (WPI), MEXT, Japan. 
A.H., Z.C. acknowledge support from the UCSB summer research mentorship program. 
A.F.Y. acknowledges support by the National Science Foundation through Enabling Quantum Leap: Convergent Accelerated Discovery Foundries for Quantum Materials Science, Engineering and Information (QAMASE-i) Award DMR-1906325; the work also made use of shared equipment sponsored under this award.

\section*{Author contributions}
L.H. and A.F.Y. conceived the project.
L.H., A.H., Z.C. developed the exfoliation technique.
Y.C. developed the van-der-Waals transfer technique.
L.H., Y.C., G.B., I.S., Y.G., T.C. and Y.J. fabricated the devices.
L.H. performed the transport and capacitance measurements as well as the data analysis.
C.Z. and W.Z. performed the nanoSQUID-on-tip measurements.
B.A.F. and A.K. fabricated the  nanoSQUID-on-tip sensors.
J.H.F. developed the numerical methods. 
J.H.F. and A.L. provided theoretical support.
T.T. and K.W. grew the hBN crystals. 
M.E.H. provided SQUID array amplifiers for nSOT readout. 
L.H. and A.F.Y. wrote the paper with inputs from all other authors.

\textbf{Data availability:}
The experimental and simulation data contained in this work is available via Zenodo at
https://doi.org/10.5281/zenodo.20753277. 

\textbf{Competing interests:}
The authors declare no competing interests.

\bibliographystyle{apsrev4-1}
\bibliography{references}

\clearpage
\newpage
\pagebreak

\section{Materials and Methods}

\textbf{Cryogenic shock exfoliation.}
For the cryogenic shock exfoliation, we use graphite crystals from NGS Naturgraphit (flaggy flakes). We have found that other graphite sources can produce subpar results. We utilize magic tape from Uline for our exfoliation and add 2 additional layers of tape for the tape to fall of in a single piece in the liquid nitrogen.
A video of the exfoliation process can be found in the supplementary materials. 
All exfoliations (including hBN) are done on silicon substrates which are \textit{not} pretreated with oxygen plasma before graphite or hBN exfoliation as the oxygen plasma increases adhesion between the flake and the chip leading to difficulties during the low-pressure transfer.

To quantify the improvement over standard mechanical exfoliation, both methods are compared on the same total Si substrate area. For the standard process, steps I and II of Fig.~\ref{fig:1}a are followed; the chip is then cooled to room temperature and the tape is peeled carefully. Flakes are screened by infrared-capable optical microscopy. Flakes with uniform areas $\geq 400\,\mu$m$^2$
and layer numbers between 3 and ${\sim}20$ are included. Rhombohedral and total areas are determined
by loops enclosing uniform stacking order, as illustrated in Fig.~\ref{figS_exfoliation_statistics}.
Regions are excluded from the analysis if they contain cracks, wrinkles, dirt, a aspect ratio with one side $\leq 5\,\mu$m, micro-domains visible in the IR image, or areas smaller than $80\,\mu$m$^2$. Since we have demonstrated that the microscopic disorder of cryogenically exfoliated flakes is comparable to that of standard flakes\cite{holleis_nanoscale_2025}, this exclusion criterion introduces approximately equal systematic error for both methods.

\textbf{Identification of crystallographic axes.}
Exfoliated graphene flakes carry a $30^\circ$ ambiguity in crystal orientation due to the bimodal distribution of armchair and zigzag facets\cite{feng_experimentally_2022}.
We resolve this using the large-scale hexagonal facets of the cleaved bulk graphite crystal
(Fig.~\ref{figS_crystal_orientation}a,b), which we identify as armchair edges\cite{kolumbus_crystallographic_2018}. 
Tracking this orientation through the exfoliation process, we assign straight flake edges either to the armchair (red hexagons) or zigzag (blue hexagons) direction, as shown in Fig.~\ref{figS_crystal_orientation}c-g.

\textbf{Preparation of transfer membrane.}
A micro-cavity in polydimethylsiloxane (PDMS) is prepared using a mold of negative $10\,\mu$m-thick SU8 photoresist. 
A $500\times250\,\mu$m rectangle is exposed, developed in SU8 developer for $1$--$2\,\mathrm{min}$, rinsed with isopropanol, and baked at $150\,^\circ$C for 5 min. 
The mold is filled with PDMS to $1$--$2\,\mathrm{mm}$ thickness, cured for 2--3 days, and cut into stamps placed with the micro-cavity facing up on a glass slide. 
Relatively thick Polycarbonate (PC) film (the thickness is confirmed by the absence of interference fringes) is stretched over the cavity, using tension. 
The polymer stack is pre-conditioned by laminating over commercial PDMS (Gel-Film from Gel-Pak) at room temperature to reduce adhesion to the silicon substrate. This step is essential and other PDMS materials might lead to subpar results.

\textbf{Van der Waals assembly process.}
Rhombohedral regions are first identified via IR imaging, the microscopically characterized via sMIM\cite{holleis_nanoscale_2025}.
Subsequently, the uniform RMG regions are isolated via AFM cutting\cite{li_ElectrodeFree_2018} and the excess graphite is removed by picking it up with hBN flakes with a straight edge which are aligned with the cuts.
Extra care is taken to stop the lamination front at the cuts in order to not contaminate the RMG surface.
Further, a generous area of $\sim 500\times 500 \mu m^2$ is cleaned around all flakes used for the vdW-stack with an STM tip or PVC polymer on a PDMS stamp.
This ensures smooth lamination during the vdW-transfer, especially for the RMG pickup.
For the vdW assembly, the PC film is first laminated over the first hBN layer at $70\,^\circ$C along the direction of the long edge of the micro-cavity. 
The stage is heated to ${\sim}125$--$130\,^\circ$C and immediately cooled to
$110\,^\circ$C, briefly exceeding the PC glass transition temperature to conform the film to the hBN topography. 
After two minutes, the hBN is picked up by delaminating the PC. 
The PC+hBN is then laminated over the RMG flake at $110\,^\circ$C; upon retraction, the lamination front is approached slowly towards the RMG and the RMG is then rapidly picked up within $\leq 1\,\mathrm{s}$ (see Fig.~\ref{figS_pressureless_pickup} and video in the supplementary materials). 
Sequential pickup of additional hBN spacers and graphite gates does not convert RMG to Bernal stacking, and standard bubble-removal steps can be applied. 
The completed stack is transferred to a substrate at $110\,^\circ$C and heated to $200\,^\circ$C to melt the PC, which is then removed in chloroform.
Lastly, the top surface is AFM-cleaned and a top graphite gate is dropped on top of the heterostructure. 
Alternatively, the top gate can be picked up with an hBN before the initial hBN dielectric and the RMG flake, however, the top graphite makes it hard to impossible to confirm rhombohedral stacking order via IR imaging post assembly.

\textbf{Transfer statistics.}
The statistics in Fig.~\ref{figS_pressureless_pickup}b were collected by seven co-authors of varying experience levels, demonstrating reproducibility of the technique. 
The low-pressure technique raises the yield from ${\sim}30\%$ to $>80\%$. Nanoscale characterization and exclusion of micro-domains\cite{holleis_nanoscale_2025} via AFM cutting techniques\cite{li_ElectrodeFree_2018} improves the yield further to $\sim90\%$. 
Transfer along crystallographic axes appears beneficial, though the relative merit of armchair versus zigzag directions remains unclear\cite{yang_stacking_2019}. 
Thicker RMG flakes seem to be more robust against relaxation. 
Further, using flakes prepared by the cryogenic shock exfoliation also does not noticeably degrade the reliability of the transfer process and any statistical change are likely due to cross-correlation with sMIM imaging and isolation of rhombohedral domains. 
The full cumulative yield reflects devices that combine all of the following steps: low-pressure transfer, sMIM characterization and AFM cutting isolation, crystallographic alignment, sequential single-stack transfer, and cryogenic shock exfoliation.

\textbf{NanoSQUID-on-tip measurement.}
Local magnetometry was performed at $T = 1.6\,\mathrm{K}$ in a liquid helium cryostat using a
nanoSQUID-on-tip (nSOT) fabricated by depositing a superconducting indium film of ${\sim}65\,\mathrm{nm}$ onto the apex of a pulled quartz tube with apex diameter ${\sim}300\,\mathrm{nm}$, following established protocols\cite{finkler_selfAligned_2010,vasyukov_scanning_2013,redekop_direct_2024}. 
The sensor achieved a field sensitivity of ${\sim}2.5\,\mathrm{nT\,Hz}^{-1/2}$, corresponding to a flux sensitivity of ${\sim}150\,n\Phi_0\,\mathrm{Hz}^{-1/2}$ ($\Phi_0 = h/2e$). 
The tip was rastered at $100\,\mathrm{nm}$ above the sample under a bias field of $B_0 \approx 7.4\,\mathrm{mT}$, while a $613.777\,\mathrm{Hz}$ square-wave voltage on the bottom gate modulated the 13-layer RMG between the spin-polarized half-metal and a neighboring non-magnetic phase\cite{patterson_superconductivity_2025}. 
The nSOT signal was read out with a series SQUID array amplifier\cite{huber_DC_2001} in a flux-locked loop, and the ac component was extracted with a lock-in amplifier. 
The out-of-plane magnetization $M(x,y)$ was determined from the measured fringe field using standard Fourier-domain techniques\cite{feldmann_resolution_2004,meltzer_direct_2017}.

\textbf{Mean free path extraction:}
To extract the temperature dependent mean free path, $\ell_{\mathrm{mfp}}(T)$, transverse magnetic focusing experiments are carried out at fixed density vs magnetic field and temperature, as shown in Fig.~\ref{fig:3}d.
From this data, the focusing peak position in field, $B_f$, its amplitude $V(T)$ and peak area $A_\mathrm{peak}(T)$ are extracted.
$B_f$ converts to the Fermi wave vector $k_F$ via
\begin{equation}
    k_F = \frac{e B_f d}{2\hbar}
\end{equation}
where $d$ is the contact separation of $\sim5 \mu m$.
Assuming a lineshape $K(B-B_f(E))$, we write the measured response as
\begin{align}
    V(B,T) \sim  \int \mathrm{d}E\, \left(- \frac{\partial f}{\partial E}\right) W(E, T) K(B -B_{f}(E)),
    \label{V_peak}
\end{align}
where $W(E,T)$ is meant to be the probability that a carrier injected at energy $E$ actually survives to reach the collecting contact and $-\partial f/\partial E $ is the derivative of the Fermi-Dirac distribution. %= \operatorname{sech}^2\!\!\left(\frac{E-E_F}{2k_BT}\right)\!\big/4k_BT
If the scattering length $\ell_{\mathrm{mfp}}$ is changing slowly over the window $k_B T \ll E_F$, then we may take $W(E,T) \approx W(E_F, T) \sim \exp(\pi d / (2 \ell_{\mathrm{mfp}}(T)))$.  
Integrating over $B$ to measure the area under the peak, one finds
\begin{align}
    A(T) \sim W(E_F,T) \int \mathrm{d}E \left(-\frac{\partial f}{\partial E}\right) \int \mathrm{d}B\, K (B - B_f(E)).
    \label{A_peak}
\end{align}
The last integral is a constant in the ballistic limit, therefore 
\begin{equation}
    A(T) \approx A_\mathrm{0} W(E_F, T),
\end{equation}
where $A_\mathrm{0}$ is the low temperature limit of the peak area.
We indeed observe a saturation of $A(T)$ below 15 K (Fig.~\ref{figS_mfp_analysis}).
Inverting and solving for $\ell_{\mathrm{mfp}}(T)$, gives
\begin{equation}
    \ell_\text{mfp}(T) = \frac{-\pi d/2}{\ln\!\left[A (T)/A_{0}\right]}
     \label{eq:lmpf}
\end{equation}
which we fit to Matthiessen's rule
\begin{equation}
    \frac{1}{\ell_\text{mfp}(T)} = \frac{1}{\ell_0}
                      + C\,T^\alpha.
\end{equation}
We find $\alpha\approx 1.56 \pm 0.05$ as well as an extrapolation of $\ell_\text{mfp}$  to $\sim$ 2$\mu m$ at 300 K.
This analysis is not well suited to determine the mean-free path at lowest temperatures, $\ell_0$, as scattering becomes exponentially suppressed once $\ell_\text{mfp}$ significantly exceeds the device size which is the reason for divergence of equation\ref{eq:lmpf}.
We can only note that $\ell_0\geq200\mu m$ below 30K; the device is thus deep in the ballistic regime at low temperatures.  

Lastly, the lineshape of equation \ref{V_peak} contains information about the band structure.
Here, we assume a Lorentzian lineshape
\begin{align}
    K(B-B_f(E))\approx\frac{\delta B_0}{\left[B_f - B_{f}(E)\right]^2 + (\delta B_0)^2},
    \label{lineshape}
\end{align}
where $\delta B_0$ is the intrinsic broadening set by the finite width of the injector and collector which describes the data reasonably well (Fig.~\ref{figS_mfp_analysis}).
Assuming parabolic bands at large $E_F$ - we get the $k_F$ and the effective mass $m^*$  dependence of equation \ref{lineshape} via $B_f - B_{f}(E) = dB_f/dE\big|_{E_F} = m^*B_n/\hbar^2k_F^2$.
The ratio of equations \ref{V_peak} and \ref{A_peak} isolates the thermal broadening
\begin{equation}
    R_\text{thermal}(T) = \int dE
        \left(-\frac{\partial f}{\partial E}\right)
        \frac{\delta B_0}{\left[m^*B_n/\hbar^2k_F^2\right]^2 + (\delta B_0)^2}.
\end{equation}
Performing a single free parameter fit to the data adjusting the effective mass, we find $m^*=0.075m_e$ at $E_F=160meV$, in excellent agreement with the calculated $m^*_\mathrm{cal}=0.078m_e$ from single particle band structure.

\textbf{Single-particle band structure.}
The single-particle band structure of rhombohedral $N$-layer graphene is computed from
a $2N \times 2N$ tight-binding Hamiltonian built on the two-site (A, B sublattice) basis
of each layer. The Hamiltonian includes five hopping parameters: the nearest-neighbor intralayer hopping $\gamma_0 = 3100\,\mathrm{meV}$; the hopping
between dimer sites $\gamma_1 = 380\,\mathrm{meV}$; the next-nearest-layer coupling
$\gamma_2 = -22\,\mathrm{meV}$; the skew interlayer hopping $\gamma_3 = -290\,\mathrm{meV}$; and the interlayer hopping between non-dimer sites
$\gamma_4 = -210\,\mathrm{meV}$. The Hamiltonian matrix elements are
\begin{align}
    H_{2l-1,\,2l}    &= -\gamma_0 f,   \quad l = 1,\ldots,N   \nonumber\\
    H_{2l,\,2l+1}    &= -\gamma_1,     \quad l = 1,\ldots,N-1 \nonumber\\
    H_{2l-1,\,2l+2}  &= -\gamma_3 f^*, \quad l = 1,\ldots,N-1 \nonumber\\
    H_{2l-1,\,2l+1}  &= -\gamma_4 f,   \quad l = 1,\ldots,N-1 \nonumber\\
    H_{2l,\,2l+2}    &= -\gamma_4 f,   \quad l = 1,\ldots,N-1 \nonumber\\
    H_{2l-1,\,2l+4}  &= -\gamma_2/2,  \quad l = 1,\ldots,N-2 
    \label{eq:hamiltonian}
\end{align}
together with their Hermitian conjugates. 
Here,
\begin{equation}
    f(\mathbf{k}) = \sum_{j=1}^{3} e^{-i\mathbf{k}\cdot\mathbf{d}_j}.
\end{equation}
No interlayer potential is included here as the relevant measurements were carried out at zero displacement field.
% The diagonal elements include three contributions. First, a uniform electrostatic potential $\Delta$ applied linearly across the stack shifts the outermost-layer sites symmetrically,
% \begin{equation}
%     H_{1,1} = H_{2,2} = +\tfrac{\Delta}{2}, \qquad
%     H_{2N-1,2N-1} = H_{2N,2N} = -\tfrac{\Delta}{2},
% \end{equation}
% with all interior diagonal elements set to zero. Second, a site energy $\delta$ is applied to the non-dimer sites of the two surface layers,
% \begin{equation}
%     H_{1,1} \mathrel{-}= \delta, \qquad H_{2N,2N} \mathrel{-}= \delta.
% \end{equation}
For the 13-layer device studied here, the Hamiltonian is diagonalized numerically on a $350\times350$ uniform grid in $(k_x, k_y)$ spanning $|k_i| \leq 0.35\,a^{-1}$, where $a = 2.46\,\text{\AA}$ is the graphene lattice constant.

The cyclotron effective mass is calculated from the thermally broadened Fermi surface
area $A(E)$ via the standard semiclassical relation
\begin{equation}
    \frac{m^*}{m_e} = \frac{\hbar^2}{2\pi m_e}
    \left|\frac{dA(E_F)}{dE}\right|,
\end{equation}
where $A(E)$ is the Fermi surface area enclosed at energy $E$.
%$E_F=160$~meV is set by the measured focusing field and resulting $k_f$.
%To account for finite temperature, $A(E)$ is thermally broadened by convolution with
%$K_T(E) = \mathrm{sech}^2(E/2k_BT)/4k_BT$ prior to differentiation.

\textbf{Numerical simulations:}  

To compare experiment with theory, we have simulated the quantum kinetic equation for an isotropic two-dimensional electron liquid within a two-rate relaxation-time approximation, following \cite{farrell_Characterizing_2026,zhang_Imaging_2026}. 
\begin{comment}
@software{fermiharmonics_zenodo,
  author = {Farrell, Jack H.},
  title = {{FermiHarmonics}: Linearized 2D Fermi-liquid Boltzmann transport on unstructured meshes},
  year = {2026},
  publisher = {Zenodo},
  doi = {10.5281/zenodo.18528662},
  url = {https://doi.org/10.5281/zenodo.18528662}
}
@article{farrellSimpleDevices2026,
  title = {Simple devices that distinguish hydrodynamic, ballistic, and diffusive transport},
  author = {Farrell, Jack H. and Lucas, Andrew},
  year = {2026},
  journal = {to appear}
}
\end{comment}

While the Fermi surface is not strictly circular in this material according to Fig.~\ref{fig:3}, we have found that the isotropic approximation, which offers key numerical advantages, qualitatively models the ballistic-hydrodynamic-diffusive crossover well.  Indeed, in the DC hydrodynamic limit, we expect corrections appear only at higher orders in the gradient expansion for a fermi surface with $\mathsf D_6$ symmetry~\cite{cook_Electron_2019}. 

Let $f(\mathbf{x},\mathbf{p})$ denote the distribution function of the electronic quasiparticles.  We write \begin{equation}
    f(\mathbf{x},\mathbf{p}) = \Theta(p_F-|\mathbf{p}|) + \delta(p_F-|\mathbf{p}|) \Phi(\mathbf{x},\theta) + \cdots 
\end{equation}
where $\Phi$ is the linearized distribution function perturbed away from equilibrium (at very low temperature, the Fermi-Dirac distribution is approximately a step function), and $\theta$ parameterizes $\mathbf{p} = (p_F\cos\theta, p_F\sin\theta)$.  The time-independent Boltzmann equation in a uniform out-of-plane magnetic field reads \begin{equation}
    (\cos\theta \partial_x+\sin\theta \partial_y) \Phi + r_{\mathrm{c}}^{-1}\partial_\theta \Phi = -\mathsf{W}[\Phi]
\end{equation}
where $r_{\mathrm{c}}$ is the cyclotron radius.
In the two-rate relaxation-time approximation, the linearized collision integral $\mathsf{W}[\Phi]$ acts as follows:  decomposing \begin{equation}
    \Phi = \frac{1}{2}A_0(\mathbf{x}) + \sum_{m=1}^\infty \left(A_m(\mathbf{x})\cos(m\theta) + B_m(\mathbf{x})\sin(m\theta)\right)
\end{equation}the linear $\mathsf{W}$ acts on each harmonic as $\mathsf{W}[\cos(m\theta)] = \ell_m^{-1} \cos(m\theta)$ (similar for sines), where \begin{equation}
    \ell_m^{-1} = \left\lbrace\begin{array}{ll} 0 &\ m=0 \\ \ell_{\mathrm{mr}}^{-1} &\ m=1 \\ \ell_{\mathrm{mr}}^{-1}+\ell_{\mathrm{ee}}^{-1} &\ m>1 \end{array}\right..
\end{equation}
$\ell_{\mathrm{mr}}$ represents the mean free path for momentum-relaxing off of (short-range) impurities, while $\ell_{\mathrm{ee}}$ represents the momentum-conserving electron-electron length.  This model is known to be a simplification of realistic scattering rates in two dimensions \cite{ledwith_hierarchy_2019} but the two fit parameters $\ell_{\mathrm{mr}}$ and $\ell_{\mathrm{ee}}$ will suffice to explain the experimental data. To solve numerically, we truncate the Fourier series at $m = M$ and employ discontinuous Galerkin spectral element methods, which are implemented in the Julia programming language through the library \texttt{Trixi.jl}~\cite{schlottke-lakemper_purely_2021, ranocha_Adaptive_2022}.

For boundary conditions, we model mixed diffuse / specular reflection from the device walls.  In the quasiparticle picture, with probability $p$, particles incident on the wall are re-emitted at a random angle back into the domain.  With probability $1-p$, the particles reflect specular about the wall's normal. We empirically choose $p=0.5$  in agreement with the measured decay as a function of $B_T$ of the magnetic focusing peak strength in the ballistic limit.  Details of how the mixed boundary condition is implemented in the Boltzmann equation picture may be found in Ref.~\cite{farrell_Characterizing_2026}.
\begin{comment}
@software{fermiharmonics_zenodo,
  author = {Farrell, Jack H.},
  title = {{FermiHarmonics}: Linearized 2D Fermi-liquid Boltzmann transport on unstructured meshes},
  year = {2026},
  publisher = {Zenodo},
  doi = {10.5281/zenodo.18528662},
  url = {https://doi.org/10.5281/zenodo.18528662}
}
@article{farrellSimpleDevices2026,
  title = {Simple devices that distinguish hydrodynamic, ballistic, and diffusive transport},
  author = {Farrell, Jack H. and Lucas, Andrew},
  year = {2026},
  journal = {to appear}
}

\end{comment}

To constrain the electron-electron and momentum scattering length, we refer to our mean-free path measurements as well as previous literature. 
$\ell_{\mathrm{mr}}$ falls into the range of 2-200$\mu$m from the TMF measurements.
$\ell_{\mathrm{ee}}$ has been recently found to reach its lower bound at the Fermi-wavelength with $\sim$50~nm in bilayer graphene\cite{zhang_Imaging_2026}.
For the spatial simulations we use this value. 
Notably, hydrodynamic effects in 13-layer graphene appear at a factor of ten higher density as in the bilayer case, hence the Fermi-wavelength is smaller and $\ell_{\mathrm{ee}}$ might be further reduced.

\textbf{Crossover from hydrodynamic to diffusive regime:} To quantify the change in scattering lengths when crossing from the hydrodynamic to the diffusive regime, we compare the spatial current flow profiles to numerical simulations.
Specifically, we use the current magnitude as a function of the $y$-coordinate at fixed $x$, as well as the current maximum along such cuts as a function of $x$.
Representative simulations are shown in Figs.~\ref{figS_hydro_diffusive}a,b, where the
white line tracks the current maximum, and panels c,d show the corresponding cuts along the $y$-direction.
With $\ell_\mathrm{ee} = 50\,\mathrm{nm}$ fixed, the current maximum shifts to lower
$y$-values and simultaneously decreases as $\ell_\mathrm{mr}$ is reduced.

Turning to the experimental nSOT data of Figs.~\ref{fig:4}c,e, we compare the spatial cuts for both density
regimes in panels f,h.
Close to charge neutrality (Fig.~\ref{figS_hydro_diffusive}e,f), the best agreement is obtained with $\ell_\mathrm{mr}$ approximately equal to the device size and small $\ell_\mathrm{ee}=50$~nm.
In the single-surface regime (Fig.~\ref{figS_hydro_diffusive}h,g), by contrast, a large $\ell_\mathrm{mr}$ matches the experiment best.
We conclude that in the latter case the system sits deep in the hydrodynamic regime,
while near charge neutrality $\ell_\mathrm{mr}$ is reduced by inelastic
electron-hole scattering.
However, $\ell_\mathrm{mr}$ never reaches the fully diffusive limit as shown in the
simulations of Fig.~\ref{figS_hydro_diffusive}a.

The current maximum along $y$ as a function of $x$ provides a complementary diagnostic.
Figures~\ref{figS_hydro_diffusive}i,j show simulations in which $\ell_\mathrm{ee}$ and
$\ell_\mathrm{mr}$ are varied, respectively, while keeping the other constant.
With $\ell_\mathrm{mr} = 200\,\mu$m fixed, increasing $\ell_\mathrm{ee}$ has little
effect on the current profile until ballistic current jets appear as nonlinearities at
large $\ell_\mathrm{ee}$.
With $\ell_\mathrm{ee} = 50\,\mathrm{nm}$ fixed, increasing $\ell_\mathrm{mr}$ shifts
the current maximum monotonically toward the boundary.
Comparing again to experiment, a reduction of $\ell_\mathrm{mr}$ to approximately the
device size provides the best match.

\clearpage
\newpage
\pagebreak

\onecolumngrid

% \begin{center}
% \textbf{\large Supplementary information }\\[5pt]
% \end{center}

\setcounter{equation}{0}
\setcounter{figure}{0}
\setcounter{table}{0}
\setcounter{section}{0}
\makeatletter
\renewcommand{\theequation}{S\arabic{equation}}
\renewcommand{\thefigure}{S\arabic{figure}}
\renewcommand{\thepage}{\arabic{page}}

% \textbf{This PDF file includes:}
% \begin{itemize}
%     \item Materials and Methods
%     \item Supplementary Figures
% \end{itemize}

\newpage
\pagebreak

\begin{figure*}
    \centering
    \includegraphics[width=\textwidth]{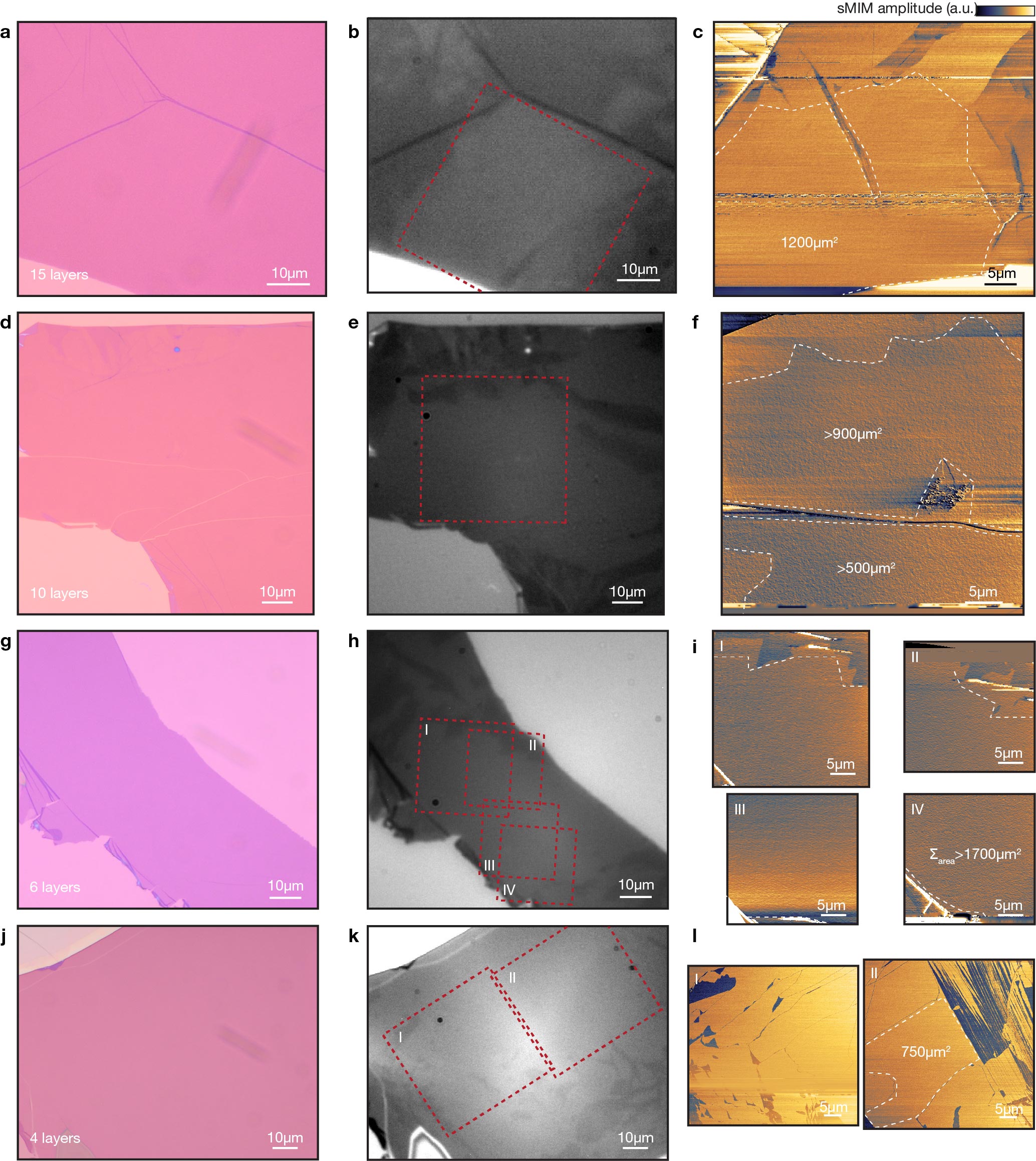}
    \caption{\textbf{Additional flakes produced by cryogenic shock exfoliation at different layer
    numbers.} Left and middle columns show optical micrographs in the visible and infrared spectra,
    respectively. The right column shows sMIM images within the region outlined by the red square
    in each IR image. White dashed lines mark regions of uniform rhombohedral stacking order.}
    \label{figS_additional_flakes}
\end{figure*}

\begin{figure*}
    \centering
    \includegraphics[width=\textwidth]{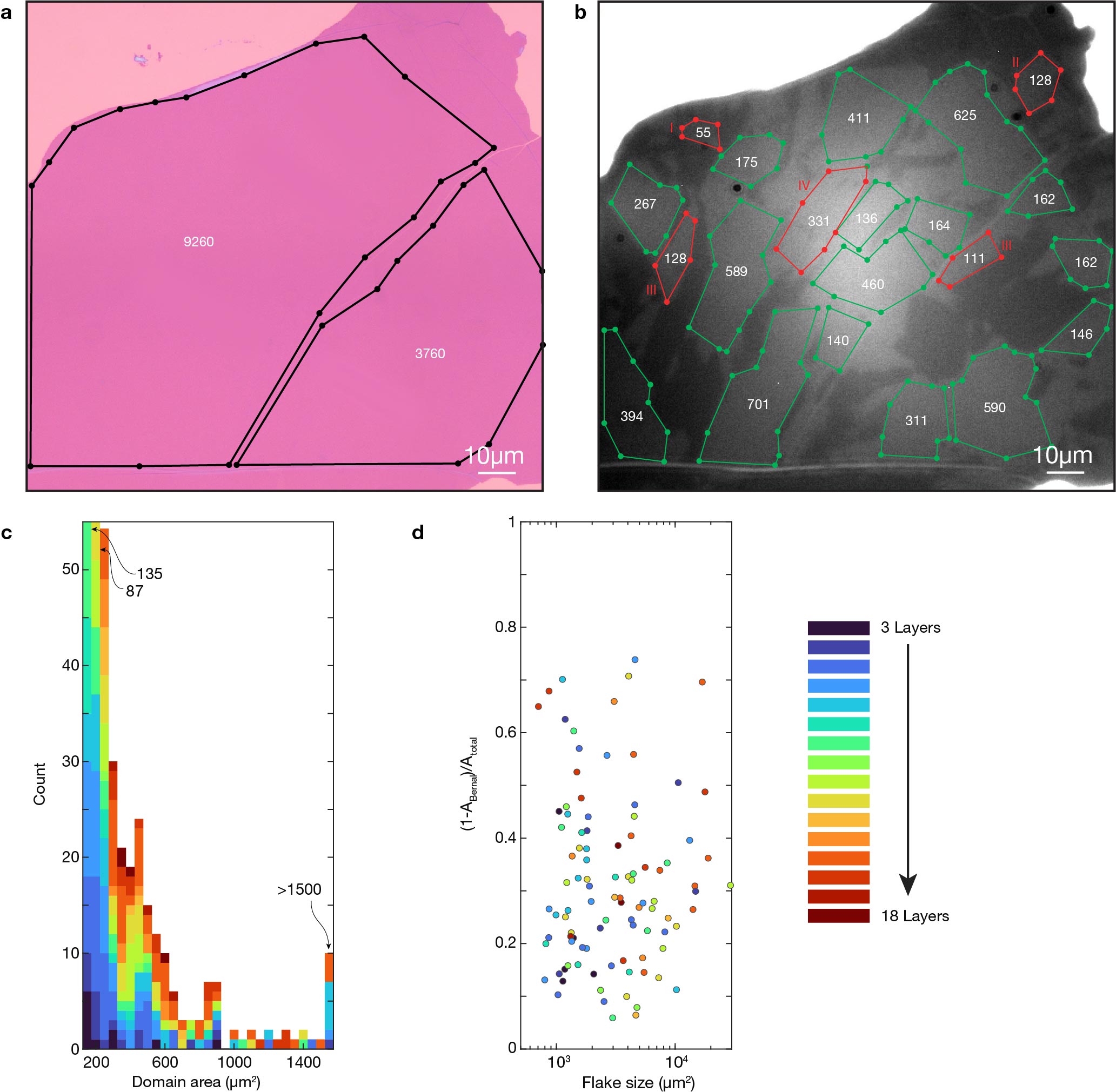}
    \caption{\textbf{Characterization of exfoliation methods.}
    \textbf{(a)} Optical micrograph in the visible spectrum (same as Fig.~\ref{fig:1}b). Black lines outline optically uniform, clean areas with the area in $\mu$m$^2$.
    \textbf{(b)} Optical micrograph in the IR spectrum (same as Fig.~\ref{fig:1}c). Loops outline bright non-Bernal regions; green loops are included in the analysis and red loops are excluded based on one of the following criteria: areas smaller than $80\,\mu$m$^2$ (I), cracks, wrinkles, dirt(II), large aspect ratio with one side $\leq 5\,\mu$m(III) or micro-domains visible in the IR image(IV).
    (c) breakdown of the histogram shown in Fig.~\ref{fig:1}h by layer number for cryogenic shock exfoliation.
    (d) distribution of flake sizes vs the non-Bernal stacking fraction for flakes produced by cryogenic shock exfoliation.}
    \label{figS_exfoliation_statistics}
\end{figure*}

\begin{figure*}
    \centering
    \includegraphics[width=\textwidth]{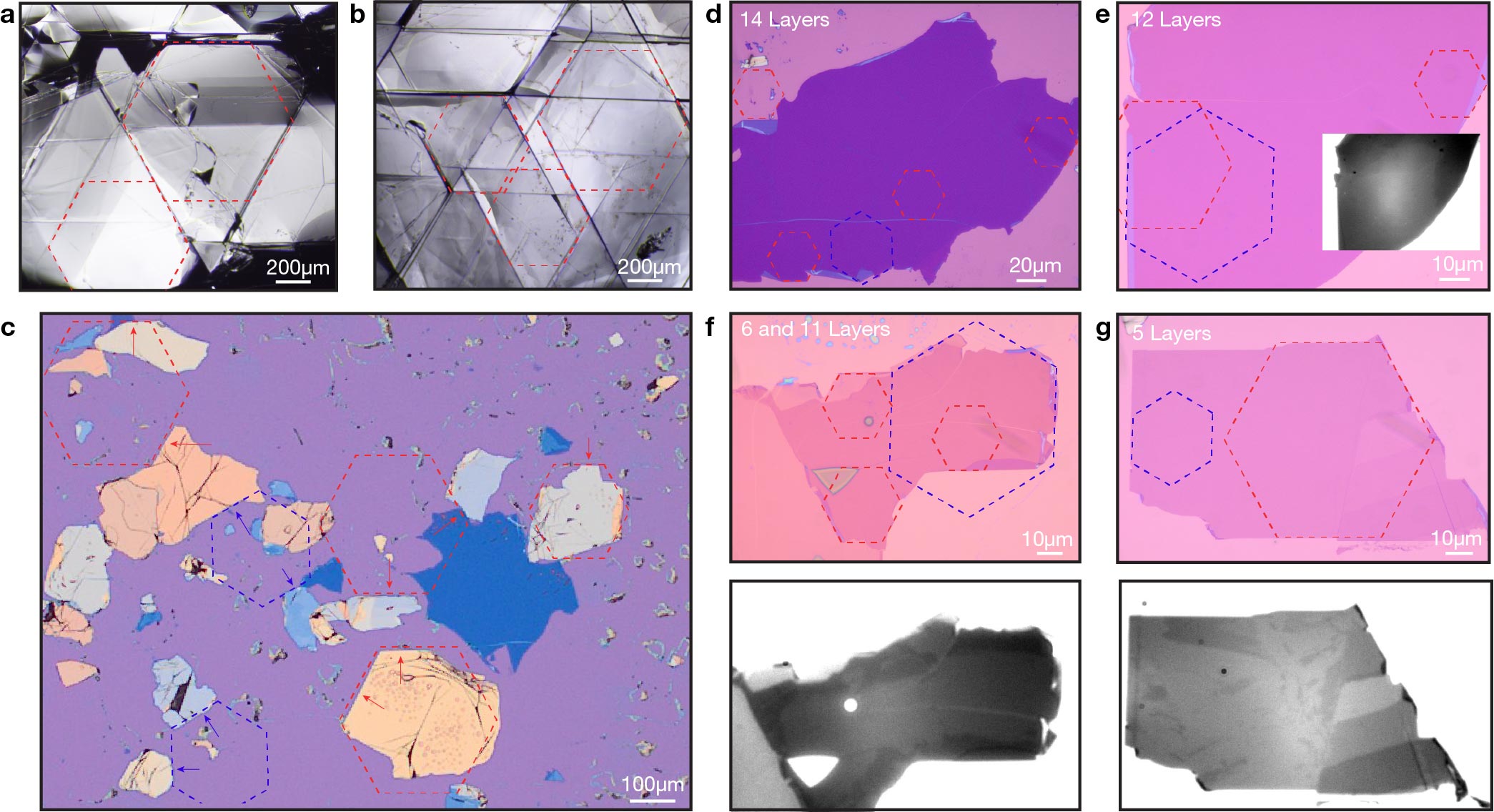}
    \caption{\textbf{Determination of crystal orientation from exfoliated flakes.}
    \textbf{(a)} Cleaved bulk graphite crystal. Red dashed hexagons outline the crystal facets.
    \textbf{(b)} Graphite exfoliated from the bulk crystal.
    \textbf{(c)} Large-scale exfoliation image. Red hexagons align with the armchair direction - tracked throughout the exfoliation process - and blue hexagons with the zigzag direction.
    \textbf{(d)} Orientational alignment for the flake in Fig.~\ref{fig:1}.
    \textbf{(e--g)} Additional examples with prominent straight edges and corresponding IR images showing large RMG areas.}
    \label{figS_crystal_orientation}
\end{figure*}

\begin{figure*}
    \centering
    \includegraphics[width=\textwidth]{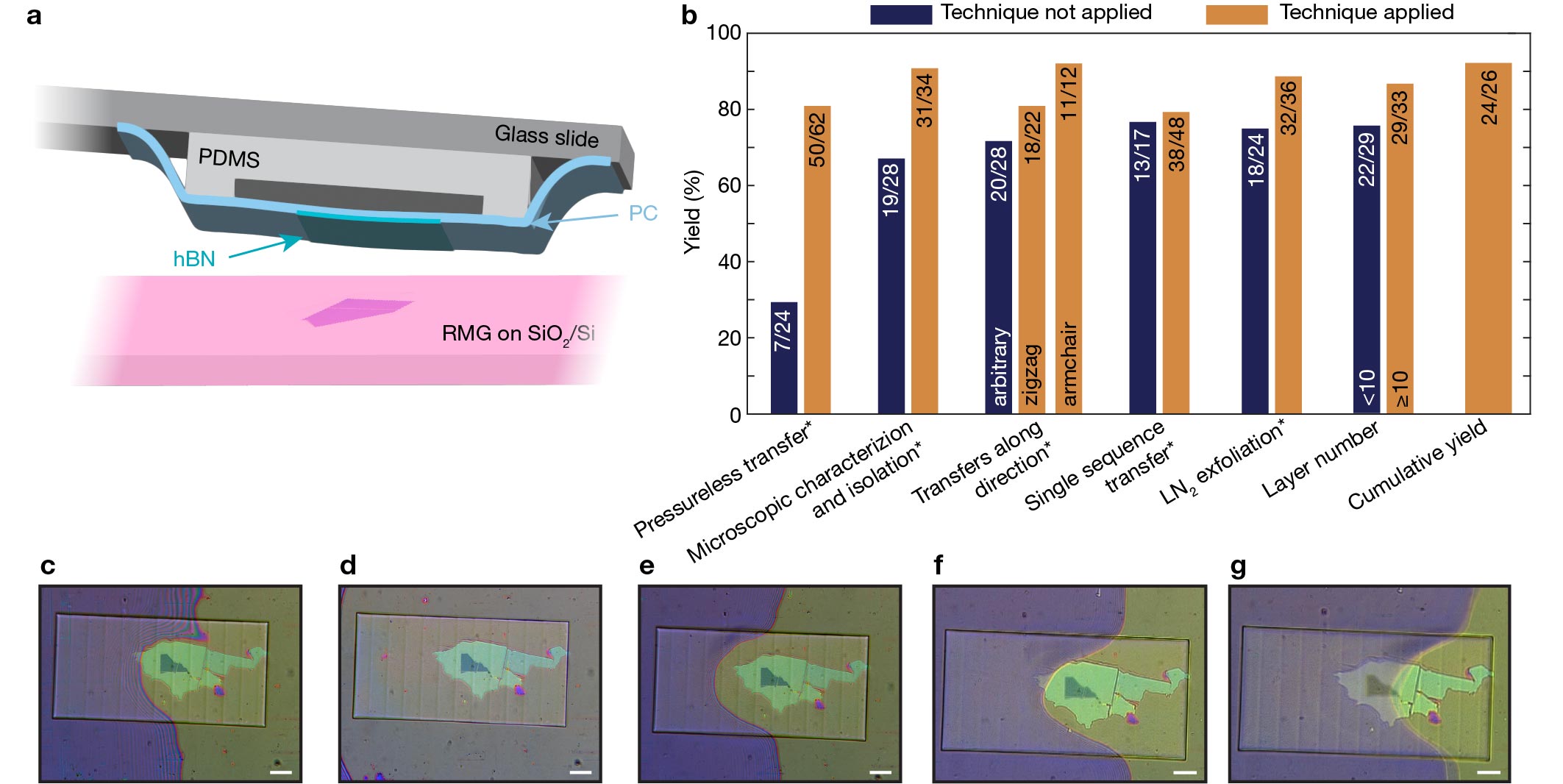}
    \caption{\textbf{Low-pressure transfer technique.}
    \textbf{(a)} Schematic of the low-pressure vdW-assembly using suspended PC membrane over a PDMS micro-cavity.
    \textbf{(b)} Transfer yield statistics. The yield is the fraction of transfers in which RMG is successfully incorporated without relaxation to Bernal stacking. The cumulative yield uses all techniques marked with $*$.
    Microscopic characterization and isolation is performed via AFM techniques\cite{holleis_nanoscale_2025,li_ElectrodeFree_2018} (see Methods).
    Single sequence transfer refers to subsequent transfer of additional dielectric layers and graphite gates after the RMG pickup.
    \textbf{(c--g)} Time-lapse of the RMG pickup process; the elapsed time between panels f and g is ${\sim}0.5$--$1\,\mathrm{s}$. The scale bar is $50\,\mu$m.}
    \label{figS_pressureless_pickup}
\end{figure*}

\begin{figure*}
    \centering
    \includegraphics[width=\textwidth]{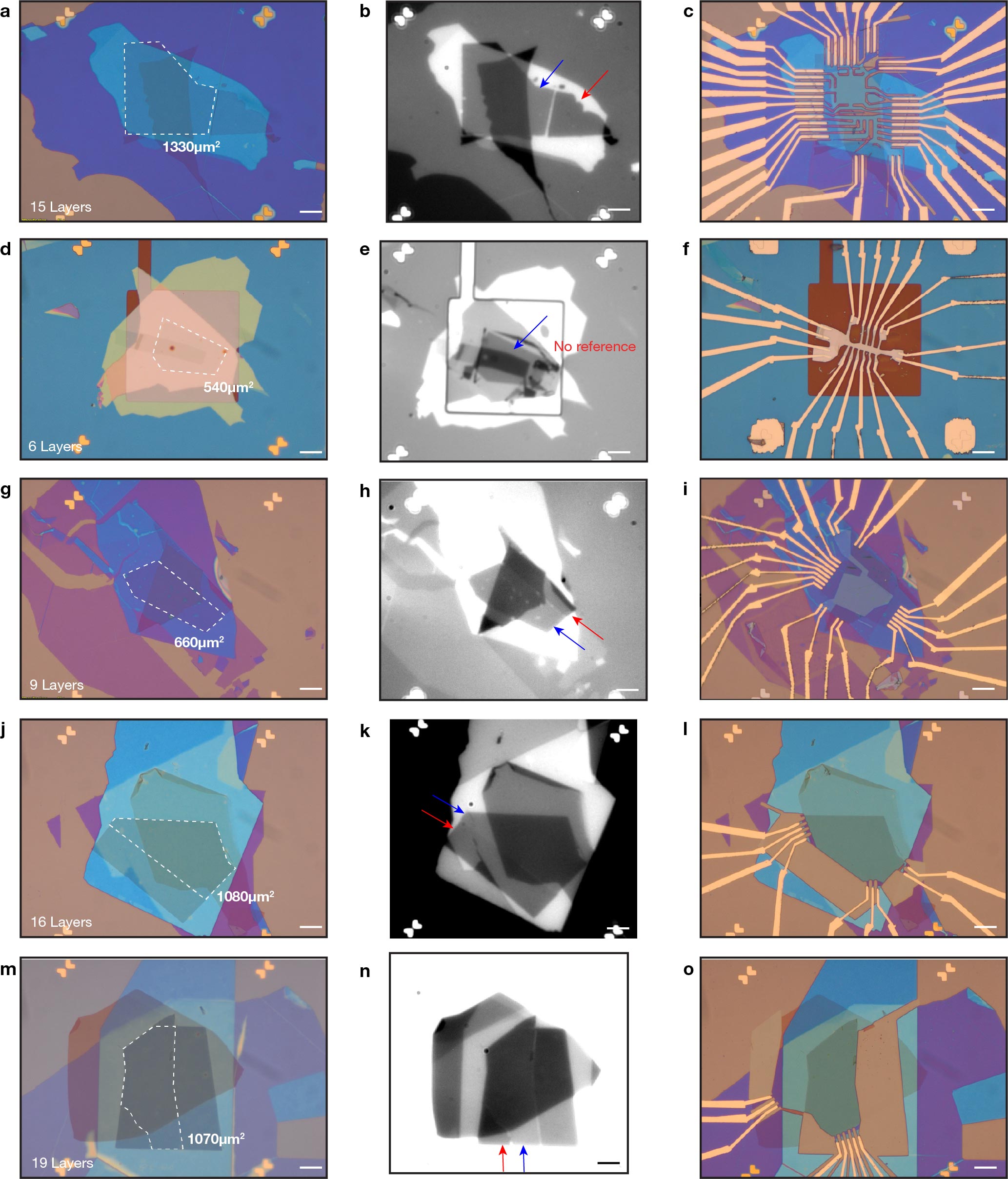}
    \caption{\textbf{Additional rhombohedral devices.} The left column shows optical micrographs after van der Waals transfer. 
    In the middle column IR images of the same devices underline the rhombohedral areas (blue arrow) and a Bernal reference region (red arrow). 
    Completed devices after nanofabrication are shown in the right column.
    Data from the device in panel d-f is shown in Fig.~\ref{figS_hexalayer}. Scale bar $10\,\mu$m in all images.}
    \label{figS_devices}
\end{figure*}

\begin{figure*}
    \centering
    \includegraphics[width=\textwidth]{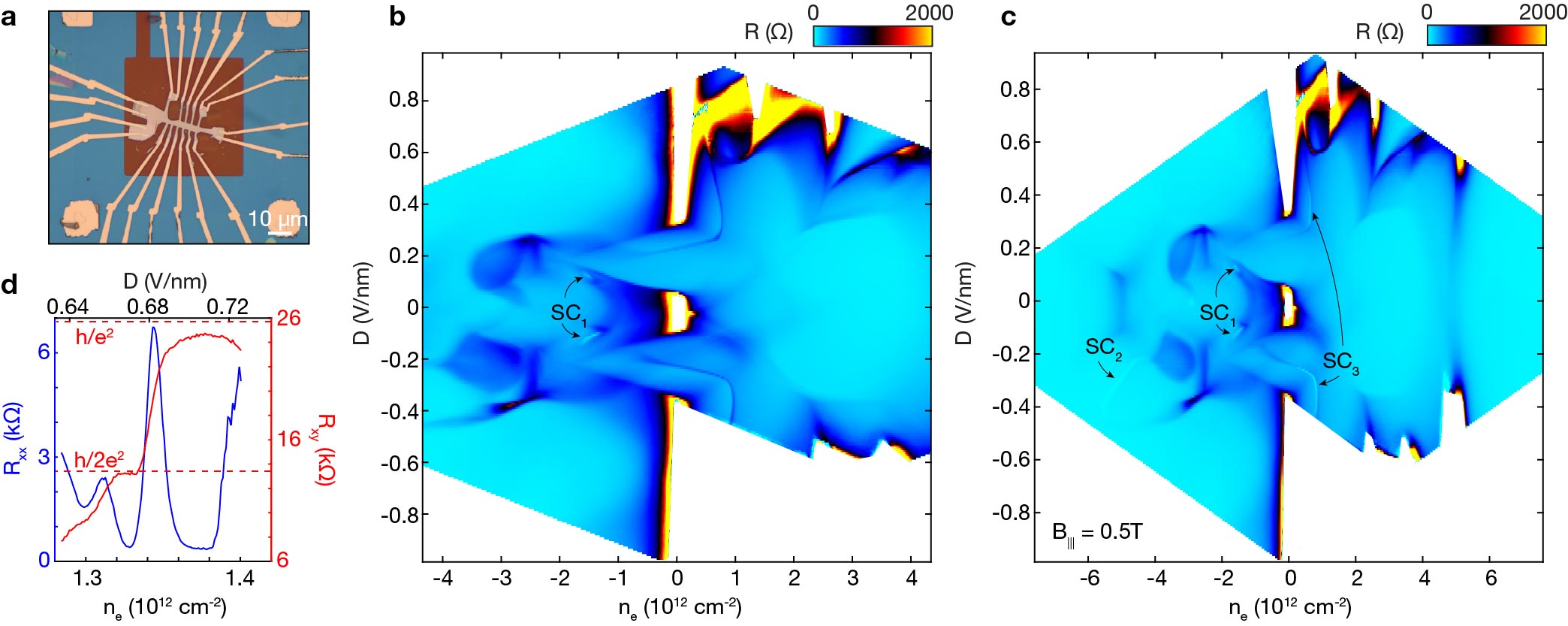}
    \caption{\textbf{Data from a rhombohedral hexalayer device aligned to hBN.} 
    \textbf{(a)} device image.
    \textbf{(b)} resistance phase diagram as function of charge carrier density $n_\mathrm{e}$ and displacement field $D$ at zero and \textbf{(c)} 0.5 T in-plane magnetic field. 
    Several superconducting pockets and symmetry broken phases appear as reported in the literature\cite{choi_superconductivity_2025,xie_MagneticFieldDriven_,morissette_evidence_2025,deng_magneticfieldinduced_2026}.
    \textbf{(d)} $R_\mathrm{xx}$ and $R_\mathrm{xy}$ measurements taken across the $\nu=1$ moir\'e peak, showing a transition from $C=2$ to $C=1$ Chern insulator. The $R_\mathrm{xx}$ ($R_\mathrm{xy}$) values are acquired by (anti-)symmetrizing data taken at $B_\mathrm{\perp} = \pm0.55$T.
    }
    \label{figS_hexalayer}
\end{figure*}

\begin{figure*}
    \centering
    \includegraphics[width=\textwidth]{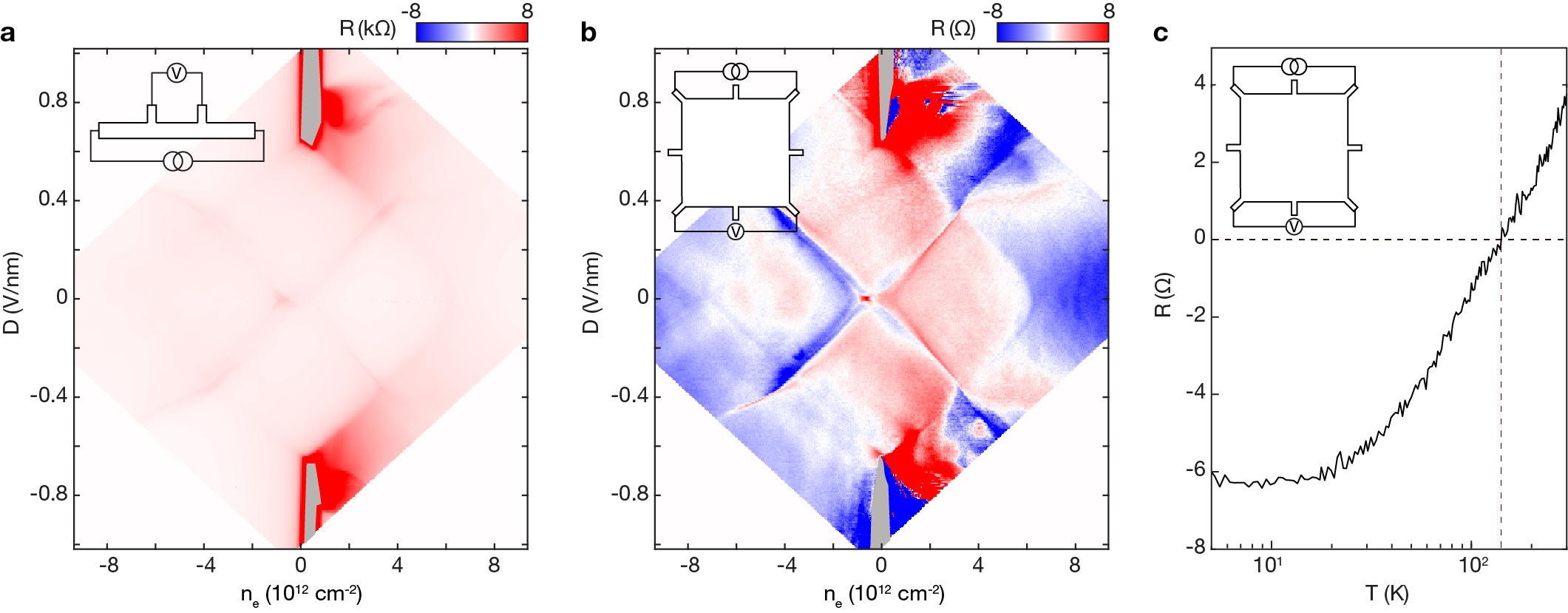}
    \caption{\textbf{Negative bend resistance due to ballistic transport.}
    \textbf{(a)} transport phase diagram measured in the narrow channel. The resistance is positive for all $n_\mathrm{e}$, $D$.
    \textbf{(b)} the same phase diagram now measured in the large square.
    Negative bend resistances are observed through large portions of the phase diagram which we ascribe to ballistic electron trajectories focusing into the negative voltage contact.
    \textbf{(c)} resistance as function of temperature at $n_\mathrm{e}=10^{13}cm^{-2}$, the same ($n_\mathrm{e},D$) point where the data in Fig.~\ref{fig:3} was taken. 
    The resistance crosses from negative to positive approximately at the point when the mean-free path has decreased to the device size.
    }
    \label{figS_negativeR}
\end{figure*}

\begin{figure*}
    \centering
    \includegraphics[width=0.7\textwidth]{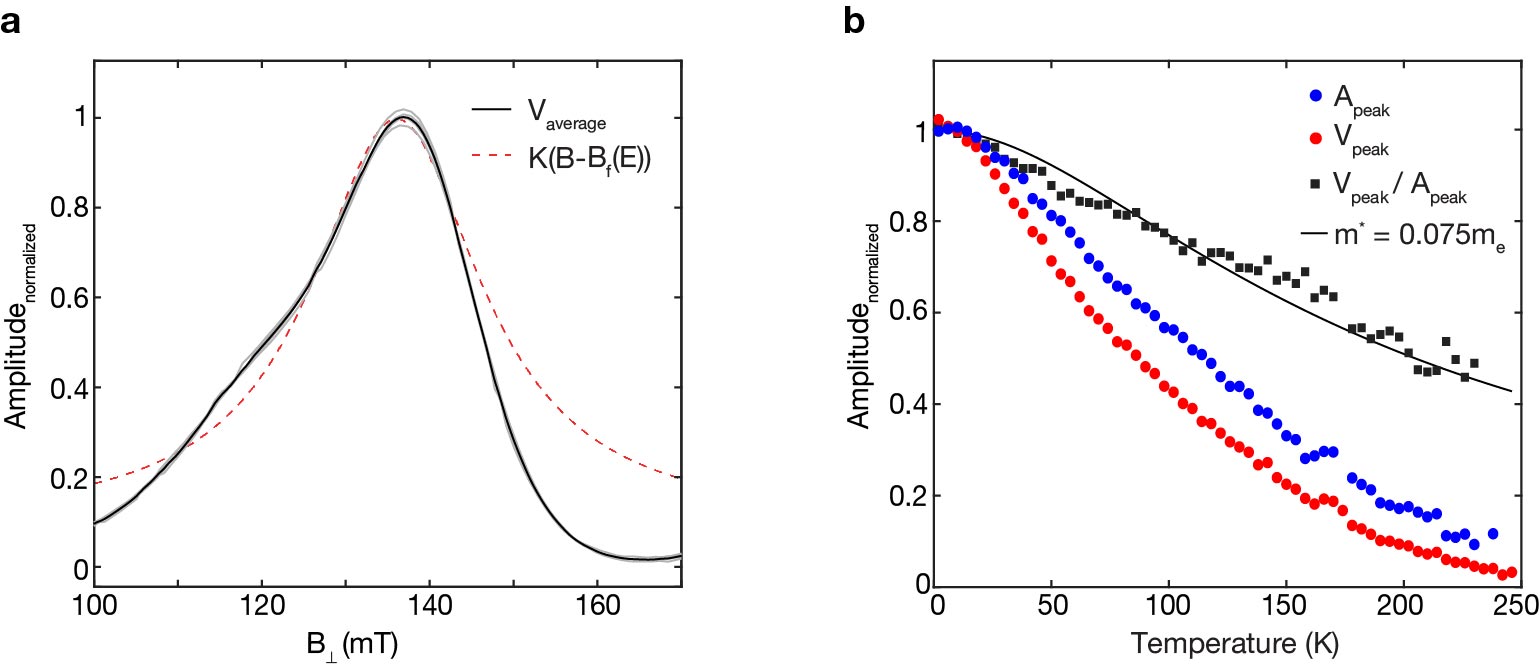}
    \caption{\textbf{Analysis of TMF data.}
    \textbf{(a)} Low-temperature datasets from 2 to 14~K (gray) and their average (black),
    illustrating near-perfect overlap. The red dashed line is a fit to the Lorentzian line shape described in the Methods.
    \textbf{(b)} Temperature-dependent amplitude normalized to the average of the four lowest
    temperature data points. 
    $A_\mathrm{peak}$ is the area under the peak; $V_\mathrm{peak}$ is the peak height at $B_f$. 
    The black line is a single-parameter fit to $R_\mathrm{thermal}$ by adjusting the effective mass.}
    \label{figS_mfp_analysis}
\end{figure*}

\begin{figure*}
    \centering
    \includegraphics[width=\textwidth]{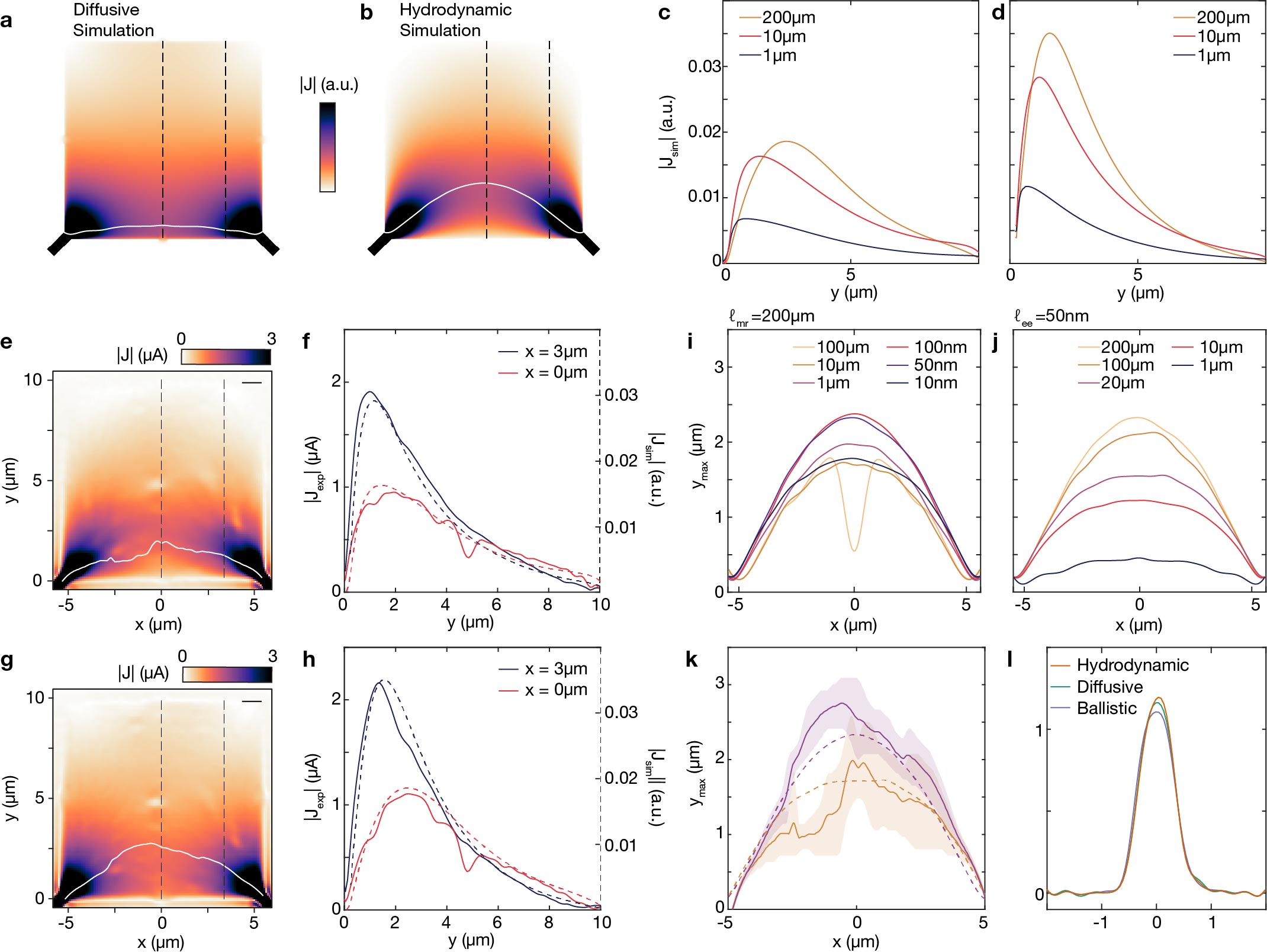}
    \caption{\textbf{Crossover from the hydrodynamic to the diffusive regime.}
    \textbf{(a,b)} Simulated current flow for diffusive ($\ell_\mathrm{mr} = 1\,\mu$m) and hydrodynamic ($\ell_\mathrm{mr} = 200\,\mu$m) regimes, with
    $\ell_\mathrm{ee} = 50\,\mathrm{nm}$ in both cases.
    Current is injected into the corners of the device, as in the experiment.
    The white line tracks the current maximum along $y$, denoted $y_\mathrm{max}$, for each value of $x$.
    \textbf{(c,d)} Simulated current magnitude as a function of $y$ at the center of the square and $3\,\mu$m from the center (dashed lines in panels a,b), with $\ell_\mathrm{ee} = 50\,\mathrm{nm}$ fixed and $\ell_\mathrm{mr}$ varied.
    \textbf{(e)} nSOT current image near charge neutrality at $n_e = 0.6\times10^{12}\,\mathrm{cm}^{-2}$, $D = 0\,\mathrm{V/nm}$, same as Fig.~\ref{fig:4}c.
    \textbf{(f)} Measured and simulated current amplitude $|J_\mathrm{exp}|$ (solid) and $|J_\mathrm{sim}|$ (dashed) along the dashed lines in panel e.
    The simulation uses $\ell_\mathrm{ee} = 50\,\mathrm{nm}$ and $\ell_\mathrm{mr} = 10\,\mu$m.
    \textbf{(g)} nSOT current image in the single-surface regime at $n_e = 6.5\times10^{12}\,\mathrm{cm}^{-2}$, $D = 0.41\,\mathrm{V/nm}$, same as Fig.~\ref{fig:4}e.
    \textbf{(h)} $|J_\mathrm{exp}|$ and $|J_\mathrm{sim}|$ for the data in panel g.
    The simulation uses $\ell_\mathrm{ee} = 50\,\mathrm{nm}$ and $\ell_\mathrm{mr} = 200\,\mu$m.
    \textbf{(i)} Simulated current maximum $y_\mathrm{max}$ along $y$ as a function of $x$, with $\ell_\mathrm{mr} = 200\,\mu$m fixed an $\ell_\mathrm{ee}$ varied.
    \textbf{(j)} Same as panel i, with $\ell_\mathrm{ee} = 50\,\mathrm{nm}$ fixed and $\ell_\mathrm{mr}$ varied.
    \textbf{(k)} Experimental $y_\mathrm{max}$ extracted from panels e (orange) and g (purple). 
    Shaded areas mark the region where the current exceeds 95\% of its maximum value. Dashed lines show the best match of simulations with
    $\ell_\mathrm{ee} = 50\,\mathrm{nm}$ and $\ell_\mathrm{mr} = 20\,\mu$m (orange) and $200\,\mu$m (purple).
    \textbf{(l)} current distribution across the narrow channel taken at the same density and displacement field as Fig.~\ref{fig:4}a,c,e. The measured current concentration agrees with our assignment of hydrodynamic, diffusive and ballistic flow analogous to the analysis performed in reference \cite{zhang_Imaging_2026}.}
    \label{figS_hydro_diffusive}
\end{figure*}

\begin{figure*}
    \centering
    \includegraphics[width=\textwidth]{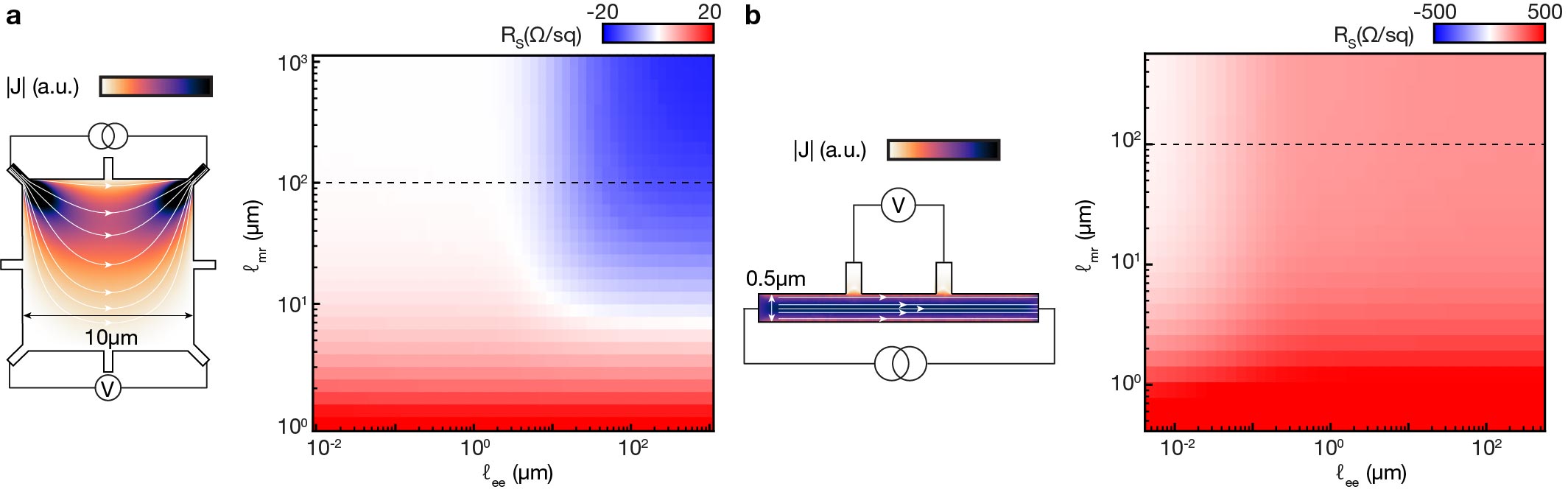}
    \caption{\textbf{Numerical simulations of resistance in the hydrodynamic regime.}
    \textbf{(a)} Resistance of the large square as a function of both $\ell_\mathrm{mr}$
    and $\ell_\mathrm{ee}$, calculated from the voltage drop across the two contacts
    divided by the total current. 
    The dashed line indicates the linecut shown in Fig.~\ref{fig:4}l. Along this cut, the resistance is small in the hydrodynamic regime (small $\ell_\mathrm{ee}$) and becomes negative as $\ell_\mathrm{ee}$ increases, due to ballistic focusing of electrons into the voltage contacts.
    \textbf{(b)} Analogous simulation for the narrow channel. Here the resistance decreases with decreasing $\ell_\mathrm{ee}$ - the Gurzhi effect.}
    \label{figS_simulations}
\end{figure*}

\end{document}